\documentclass[onecolumn,aps,pra,preprint,amsmath,amssymb,groupedaddress,superscriptaddress]{revtex4-1}

\usepackage{graphicx}
\usepackage{graphics}
\usepackage{dcolumn}
\usepackage{bm}
\usepackage{color}
\usepackage{soul}
\usepackage{textcomp}
\usepackage{soul}
\usepackage{rotating}
\usepackage{setspace}
\usepackage{multirow}
\usepackage{amssymb}
\usepackage{textcomp}
\usepackage{mathrsfs}
\usepackage[mathlines]{lineno}
\usepackage{amsmath}

\usepackage{microtype}
\usepackage[hyperindex=true,pdftitle={Molecular Excitation by Blackbody Radiation},
colorlinks=true,pagebackref=false,citecolor=blue,plainpages=false,pdfpagelabels,
linkcolor=blue,urlcolor=blue]{hyperref}

\def\proj#1#2{\left| #1 \left\rangle \right\langle #2\right|}

\def\ket#1{|\,#1\, \rangle}

\def\Xint#1{\mathchoice
{\XXint\displaystyle\textstyle{#1}}%
{\XXint\textstyle\scriptstyle{#1}}%
{\XXint\scriptstyle\scriptscriptstyle{#1}}%
{\XXint\scriptscriptstyle\scriptscriptstyle{#1}}%
\!\int}
\def\XXint#1#2#3{{\setbox0=\hbox{$#1{#2#3}{\int}$}
\vcenter{\hbox{$#2#3$}}\kern-.5\wd0}}

\def\dashint{\Xint-}

\begin{document}
\title{
Open System Perspective on Incoherent Excitation of Light Harvesting Systems
}
\author{Leonardo A. Pach\'on}
\affiliation{Grupo de F\'isica At\'omica y Molecular, Instituto de F\'{\i}sica,
Facultad de Ciencias Exactas y Naturales,
Universidad de Antioquia UdeA; Calle 70 No. 52-21, Medell\'in, Colombia.}
\author{Juan D. Botero}
\affiliation{Grupo de F\'isica At\'omica y Molecular, Instituto de F\'{\i}sica,
Facultad de Ciencias Exactas y Naturales,
Universidad de Antioquia UdeA; Calle 70 No. 52-21, Medell\'in, Colombia.}
%
%
\author{Paul Brumer}		   
\affiliation{Chemical Physics Theory Group, Department of Chemistry and
Center for Quantum Information and Quantum Control,
\\ University of Toronto, Toronto, Canada M5S 3H6}
%
%
\begin{abstract}
The nature of excited states of open quantum systems produced by incoherent natural thermal
light is analyzed based on a description of the quantum dynamical map.
Natural thermal light is shown to generate long-lasting coherent dynamics because of (i) 
the super-Ohmic character of the radiation, and (ii) the absence of pure dephasing dynamics.
In the presence of an environment,  the long-lasting coherences induced
by  suddenly turned-on incoherent light dissipate and stationary coherences are established.
As a particular application, dynamics in a subunit of the PC-645 light-harvesting complex is considered
where it is further shown that aspects of the energy pathways landscape depend
on  the nature of the exciting light and number of chromophores
excited.  Specifically,  pulsed  laser  and  natural  broadband
incoherent  excitation  induce  significantly  different energy
transfer pathways. In addition, we discuss differences in perspective
associated with the eigenstate vs site basis, and note an important difference
in  the  phase  of  system coherences when coupled to blackbody
radiation or when coupled to a phonon background. Finally, an Appendix
contains an open systems example of the loss of coherence as the turn on time
of the light assumes natural time scales.
\end{abstract}

\date{\today}


\maketitle

\section{Introduction}
Interest in light-induced biological processes, such as photosynthesis and vision,
as well as  in the design of energy efficient photovoltaic systems, and general aspects
of noise-induced coherence has renewed interest in excitation with incoherent (e.g. solar)
radiation.
Many of these processes have been studied with coherent laser excitation \cite{EC&07,CW&10},
but we, and others, have convincingly demonstrated that the dynamical response of molecular systems
to coherent vs incoherent radiation is dramatically different  \cite{JB91,MV10,BS12}.
Specifically, even in the case of rapid turn on of the incoherent source, which produces noise
induced Fano coherences, the steady state that results after a short time shows no light-induced
coherences.
This is distinctly different from the results of pulsed laser excitation, as
most recently shown for coherent vs. incoherent excitation with the same spectrum \cite{Aurelia}.
Thus, the coherent pulsed laser  experiments provide important insight into the nature of the
system Hamiltonian and of the coupling of the system to the environment, but describe coherences in
dynamics that is dramatically different than natural processes
\cite{MV10,BS12,FOS12,PB13,KYR13,HSB13,PB14a,CMM14,CB&14,SB14,PC&14}.

Given the significance of the incoherent excitation process, and the difficulty of doing
experiments with incoherent light, theoretical/computational studies become increasingly
important.
As a result, a number of such computations have been carried out  \cite{JB91,MV10,HSB13,PB14a,SB14},
and insight into dynamics induced by incoherent radiation have been obtained.
However, a number of significant issues, addressed in this paper, have not been explored.
These include the origin of the extraordinarily long decoherence times upon blackbody excitation
and primary dependence of decoherence times on system level spacings \cite{PB11,SB14,TB14},
differences in the decoherence arising from blackbody radiation vs. a phonon bath,
differences in energy transfer \textit{pathways} that can result from laser vs. in-vivo excitation,
and the effect of simultaneous excitation of multiple chromophores in light harvesting systems.
These issues, of relevance to a wide variety of processes, are addressed in this paper, using
model photosynthetic light harvesting systems as examples.

Specifically, this paper is organized as follows: Section II develops the molecular
interaction with blackbody radiation from the viewpoint of open system quantum mechanics.
Emphasis is placed on the matrix elements of the quantum dynamical map, which allows
insight for a wide variety of conditions.
The significance of the resultant super-ohmic spectral density is emphasized and the
differences between system-radiation field and system-phonon bath interaction is noted.
Dynamics results for model systems are provided in Section III, starting with a
single chromophore, followed by a model PC645 dimer and the four chromophore PC645 system.
Interaction of the system with an incoherent radiative bath and a phonon bath are
independently characterized, as is simultaneous interaction with both. Coherences in
both the eigenstate and site basis are discussed. A previously proposed model for
excitation with incoherent light is analyzed and shown wanting, emphasizing the
significance of properly carrying out the full excitation step. Section IV summarizes
the contributions in this paper to general theory of light induced processes in
incoherent light, providing new insights from an open systems perspective.

Note that the dynamics considered in this paper assumes the instantaneous turn-on of the
blackbody radiation. As a consequence, light-induced coherences shown below
arise primarily from the rapid turn-on, and the focus here, as in many other studies,
is how these coherences disappear upon interactions
with the radiative or phonon environment. However, such light-induced coherences are not
expected to occur in natural systems where the turn-on time is slow in comparison
with molecular time scales\cite{DTB16b}. This issue is discussed
in Appendix A within an open quantum systems perspective.

\section{Light-harvesting antenna systems under sunlight illumination}
\subsection{The System}

To model electronic energy transfer in a light-harvesting antenna under sunlight illumination,
the electronic degrees of freedom of the molecular aggregate are considered as the system
of interest and the protein and solvent environment are treated as a local phonon bath.
Sunlight, properly described as thermal radiation,  is allowed to excite the entire aggregate.
The total Hamiltonian for an $N$-site systems in the site basis $|\epsilon_j\rangle$ is then given by
\begin{equation}
\label{equ:hamiltonian}
\begin{split}
\hat{H} &= \sum_{j=1}^N \left(E_{g_j} \hat{1}_j + \epsilon_j \proj{\epsilon_j}{\epsilon_j}\right)
- \sum_{j} \hat{\boldsymbol{\mu}}_j\cdot \hat{\mathbf{E}}(t)
+
\sum_{j} \sum_{l} \hbar g_{jl} \proj{\epsilon_j}{\epsilon_j}\left( \hat{b}^{(j)}_{l}
+{\hat{b}_{l}}^{(j)\dag}  \right)
\\ &+
\frac{\hbar}{2} \sum_{j,k} \textrm{D}_{jk} \left( \sigma_x^{(j)} \sigma_x^{(k)} + \sigma_y^{(j)} \sigma_y^{(k)}\right)
+
\sum_{\mathbf{k},s} \hbar c k \hat{a}_{\mathbf{k},s}^{\dag} \hat{a}_{\mathbf{k},s}
+ \sum_j^N \sum_{l} \hbar \omega_l^{(j)} \hat{b}_{l}^{(j)\dag} \hat{b}_{l}^{(j)},
\end{split}
\end{equation}
with $\epsilon_j$ being the electronic site energy of state $|\epsilon_j\rangle$, $E_{g_j}$ is the ground state
energy and $\textrm{D}_{jk}$ denoting the electronic couplings between the $j^\mathrm{th}$ and the
$k^\mathrm{th}$
chromophore.
Here, $\hat{b}_l^{(j)\dag}$ ($\hat{b}_l^{(j)}$) is the creation (annihilation) operator of a $l^\mathrm{th}$
phonon mode of frequency $\omega_l^{(j)}$ which is in contact with the $j^\mathrm{th}$ chromophore.
Similarly with $\hat{a}_{\mathbf{k},s}^{\dag}$ ($\hat{a}_{\mathbf{k},s} $) for the field modes.
$\hat{\mathbf{E}}(t)$ denotes the electric field of the radiation \cite{MW95} and is given by
$ \hat{\mathbf{E}}(t)= \hat{\mathbf{E}}^{(+)}(t)+ \hat{\mathbf{E}}^{(-)}(t)$ with
$
 \hat{\mathbf{E}}^{(+)}(t)=
 \mathrm{i} \sum_{\mathbf{k},s} \left(\frac{\hbar \omega}{2 \epsilon_0 V}\right)^{1/2}
 \hat{a}_{\mathbf{k},s} (\varepsilon_{\mathbf{k},s})e^{-\mathrm{i} \omega t}
$
and $ \hat{\mathbf{E}}^{(-)}(t)=\left[\hat{\mathbf{E}}^{(+)}(t)\right]^\dagger$.
Note that the effect of the radiation on the environment is neglected since it is assumed to
carry negligible oscillator strength.

\subsection{System Dynamics}
To obtain the dynamics of the electronic degrees of freedom in the presence of the incoherent
radiation and vibrational/phonon bath, we solve the density matrix dynamics
in the system eigenstate basis $\{|e\rangle\}$. The standard master equation
for the reduced density matrix $\rho$ derived for thermal
baths comprised of harmonic modes (cf. Chapter 3 in Ref.~\citenum{MK01}),  is
\begin{equation}
\label{equ:RFnMME}
\dot{\rho}_{ab} = - \mathrm{i}\omega_{ab} \rho_{ab} -
\sum_{c,d}\left[\mathsf{R}^{\mathrm{tb}}_{ab,cd} + \mathsf{R}^{\mathrm{bb}}_{ab,cd}\right]\rho_{cd}.
\end{equation}
Here $\mathsf{R}^{\mathrm{tb}}_{ab,cd} $ and $\mathsf{R}^{\mathrm{bb}}_{ab,cd} $ account for the
non-unitary contribution to the dynamics by the thermal bath (tb) and by the sunlight, described as
blackbody radiation (bb).
These are given by
\begin{equation}
\label{equ:Rtensor}
\mathsf{R}^{\mathrm{tb,bb}}_{ab,cd}(t) = \delta_{ac} \sum_e \Gamma^{\mathrm{tb,bb}}_{be,ed}(\omega_{de})
                     + \delta_{bd} \sum_e \Gamma^{\mathrm{tb,bb}}_{ae,ec}(\omega_{ce})
                      - \Gamma^{\mathrm{tb,bb}}_{ca,bd}(\omega_{db})
                      - \Gamma^{\mathrm{tb,bb}}_{db,ac}(\omega_{ca}),
\end{equation}
with
\begin{equation}
\begin{split}
\Gamma^{\mathrm{tb,bb}}_{ab,cd}(\omega) =
\frac{1}{2}\int\limits_{-\infty}^{\infty} \mathrm{d}\tau
\mathsf{M}^{\mathrm{tb,bb}}_{ab,cd}(\tau)\mathrm{e}^{\mathrm{i}\omega_{dc}\tau}
+
\frac{1}{2}\int\limits_0^{\infty} \mathrm{d}\tau
\left[
\mathsf{M}^{\mathrm{tb,bb}}_{ab,cd}(\tau)
\mathrm{e}^{\mathrm{i}\omega_{dc}\tau}
- \mathsf{M}^{*\mathrm{tb,bb}}_{ab,cd}(\tau)\mathrm{e}^{\mathrm{-i}\omega_{dc}\tau}
\right].
\end{split}
\end{equation}
The real part of $\Gamma^{\mathrm{tb,bb}}_{ab,cd}$ describes an irreversible redistribution
of the amplitudes contained in the various parts of reduced density matrix.
The imaginary part introduces terms that can be interpreted as a modification of the transition
frequencies and of the respective mean-field matrix elements.
Here $\mathsf{M}^{\mathrm{tb,bb}}_{ab,cd}(t)$, the memory matrix elements (cf. Chapter~3 in
Ref.~\citenum{MK01}),  determine the time span for correlations.

To define these elements more clearly, we
 denote the observables of the electronic
system that are coupled to the environment by $\hat{K}^{\mathrm{tb}}_u$, and the observables
of the environment that are coupled to the electronic system by $\hat{\Phi}^{\mathrm{tb}}_u$.
Thus, the interaction term with the thermal bath can be written as
$\sum_u \hat{\Phi}^{\mathrm{tb}}_u\otimes \hat{K}^{\mathrm{tb}}_u$, with a similar form for
the coupling to the electromagnetic radiation.
This representation allows us to cast the memory matrix elements as
$\mathsf{M}^{\mathrm{tb,bb}}_{ab,cd}(t) = \sum_{u,v}C^{\mathrm{tb,bb}}_{uv}(t)
K^{\mathrm{tb,bb}}_{u,ab} K^{\mathrm{tb,bb}}_{v,cd}$,
with $K^{\mathrm{tb,bb}}_{u,ab} =  \langle a | \hat{K}^{\mathrm{tb,bb}}_u | b \rangle$. Here,
the reservoir correlation function, $C^{\mathrm{tb,bb}}_{uv}(t)$, is given by
$
C^{\mathrm{tb,bb}}_{uv}(t) = \frac{1}{\hbar^2}\langle \Phi^{\mathrm{tb,bb}}_u(t)
\Phi^{\mathrm{tb,bb}}_v(0)\rangle_{\mathrm{tb,bb}},
$
where $\langle \Phi^{\mathrm{tb,bb}}_u(t) \Phi^{\mathrm{tb,bb}}_v(0)\rangle_{\mathrm{tb,bb}} =
\mathrm{tr}[\hat{\rho}^{\mathrm{tb,bb}}_{\mathrm{equ}} \Phi^{\mathrm{tb,bb}}_u(t) \Phi^{\mathrm{tb,bb}}_v(0)]$
and $\langle\Phi^{\mathrm{tb,bb}}_u\rangle_{\mathrm{tb,bb}} =0$ is assumed.
Note that Eq. (\ref{equ:RFnMME}) is then a second-order, non-secular master equation that
incorporates the structure of the environment through $\Gamma^{\mathrm{tb,bb}}_{ab,cd}(\omega)$
(see Ref.~\cite{PYB13}  for a completely general, second order, non-secular, non-Markovian master
equation).

As is standard, the correlation function for each environment vibrational/phonon mode
in Eq.~(\ref{equ:hamiltonian}) is given by
$
C^{\mathrm{tb}}_j(t) = \int_0^\infty  \mathrm{d} \omega \omega^2 J^{\mathrm{tb}}_j(\omega)
\left[\coth\left(\frac{1}{2} \hbar \omega \beta \right)\cos(\omega t)- \mathrm{i}\sin(\omega t)\right]
$
where the characteristics of the local environment in the $j^\mathrm{th}$ chromophore
are condensed in the spectral density  
\begin{equation}
\label{equ:spcfcJ}
\omega^2 J^{\mathrm{tb}}_j(\omega) = 2 \lambda_j \Lambda_j \omega/\hbar(\omega^2 + \lambda_j^2).
\end{equation}
For the simulations  below, identical spectral densities on each site are taken, with
$\lambda = 100$~cm$^{-1}$ and with various values of the reorganization energy $\Lambda$.  Note that here $C_{uv}$ is replaced by 
$C_j$ since it denotes a single observable coupled to the bath.  (As an aside, 
we note that spectral densities may be directly determined from experiment
as discussed in Ref. \cite{spectraldensities}.)

Consider now the system-blackbody radiative interaction.
Despite the fact that the spectral properties of blackbody radiation have been discussed
from the perspective of open quantum systems \cite{FLO85,PB13,PB14}, the significant 
aspects of this formulation has been overlooked in the considerations on the nature of states 
prepared and sustained by incoherent light.
Hence, we take pains to obtain the spectral density for blackbody radiation in detail, with
enlightening results.

\subsection{Open-quantum-system description of the incoherent light}
A cartesian component of the electric field at $\mathbf{r}=0$ (dipole approximation)
can be written as
%
$ \hat{\mathbf{E}}_i(t)= \hat{\mathbf{E}}_i^{(+)}(t)+ \hat{\mathbf{E}}_i^{(-)}(t)$ with $i=\{x,y,z\}$
where
$
 \hat{\mathbf{E}}_i^{(+)}(t)=
 \mathrm{i} \sum_{\mathbf{k},s} \left(\frac{\hbar \omega}{2 \epsilon_0 V}\right)^{1/2}
 \hat{a}_{\mathbf{k},s} (\varepsilon_{\mathbf{k},s})_i e^{-\mathrm{i} \omega t}
$
and $ \hat{\mathbf{E}}_i^{(-)}(t)=\left[\hat{\mathbf{E}}_i^{(+)}(t)\right]^\dagger$.
Hence, the two point correlation function of the electric field
$\langle \hat{\mathbf{E}}_i(t) \hat{\mathbf{E}}_j(0)\rangle$ comprises four terms
\begin{equation}
 \label{equ:2point}
 \begin{split}
 \langle \hat{\mathbf{E}}_i(t)\mathbf{E}_j(0)\rangle&=
 \langle \hat{\mathbf{E}}_i^{(-)}(t) \hat{\mathbf{E}}_j^{(+)}(0)\rangle+
 \langle \hat{\mathbf{E}}_i^{(+)}(t) \hat{\mathbf{E}}_j^{(-)}(0)\rangle
 \\
 &+\langle \hat{\mathbf{E}}_i^{(+)}(t) \hat{\mathbf{E}}_j^{(+)}(0)\rangle+
 \langle \hat{\mathbf{E}}_i^{(-)}(t) \hat{\mathbf{E}}_j^{(-)}(0)\rangle.
 \end{split}
\end{equation}
With the average value of the number operator for the incoherent light in a
thermal distribution given by
$ \langle a^\dagger_{\mathbf{k},s} a_{\mathbf{k'},s'}\rangle=
\frac{\delta_{\mathbf{k},\mathbf{k'}}\delta_{s,s'}}{e^{\hbar \omega/k_\mathrm{B} T}-1}$ and that
$\sum_s (\varepsilon^*_{\mathbf{k},s})_i(\varepsilon_{\mathbf{k},s})_j = \delta_{i,j}-k_i k_j/k^2 $,
the first term reads
$
 \langle\hat{\mathbf{E}}_i^{(-)}(t)\hat{\mathbf{E}}_j^{(+)}(0)\rangle=
 \sum_{\mathbf{k}} \left(\frac{\hbar \omega}{2 \epsilon_0 V}\right)\left(\delta_{i,j}-\frac{k_i k_j}{k^2 }\right)
 \frac{e^{\mathrm{i} \omega t}}{e^{\hbar \omega/k_\mathrm{B} T}-1}.
$
Taking the continuum limit
$\sum_\mathbf{k} \rightarrow \frac{V}{(2\pi)^3}\int \mathrm{d}\mathbf{k}$,
and using spherical coordinates
$\mathrm{d}\mathbf{k}=k^2 \mathrm{d}k\mathrm{d}\Omega_k
=c^{-3}\omega^2\mathrm{d}\omega\mathrm{d}\Omega_\omega$, gives
the final expression for the first term as
\begin{equation}
\label{equ:2point-+}
\begin{split}
  \langle \hat{\mathbf{E}}_i^{(-)}(t) \hat{\mathbf{E}}_j^{(+)}(0)\rangle &=
  \frac{\hbar}{4\epsilon_0\pi^2c^3}\int\limits_0^\infty  \mathrm{d}\omega
    \left(\delta_{ij}-\frac{k_i k_j}{k^2 }\right)
 \left[\frac{\omega^3}{e^{\hbar \omega/k_{\mathrm B} T}-1}\right] .
  \end{split}
\end{equation}
The second term of the right hand side of the Eq.~(\ref{equ:2point})
is obtained by using the commutation relation between the creation and annihilation operators
and is given by
$
  \langle \hat{\mathbf{E}}_i^{(+)}(t) \hat{\mathbf{E}}_j^{(-)}(0)\rangle =
  \frac{\hbar}{4\epsilon_0\pi^2c^3}\int\limits_0^\infty  \mathrm{d}\omega
    \left(\delta_{ij}-\frac{k_i k_j}{k^2 }\right)
 \left[\frac{\omega^3}{e^{\hbar \omega/k_{\mathrm B} T}-1}+\omega^3\right] .
$
Finally, the last two terms in equation (\ref{equ:2point}) are equal to zero since
$\langle a^\dagger_{\mathbf{k},s} a^\dagger_{\mathbf{k'},s'}\rangle=
\langle a_{\mathbf{k},s} a_{\mathbf{k'},s'}\rangle=0$.
Thus, the two point correlation function of the incoherent light reads
 \begin{equation}
  \langle\hat{\mathbf{E}}_i(t)\hat{\mathbf{E}}_j(0)\rangle =
  \frac{\hbar}{4\epsilon_0\pi^2c^3}
  \int\limits_0^\infty  \mathrm{d}\omega
\left(\delta_{ij}-\frac{k_i k_j}{k^2 }\right)  \omega^3
 \left[\coth\left(\frac{\hbar \omega}{2k_{\mathrm{B}}T } \right)\cos(\omega t)- \mathrm{i}\sin(\omega t)\right].
\end{equation}
By assuming  homogeneous polarization with only two of the three components
of $\mathbf{k}$ contributing to the coupling in each spatial direction (transversality condition),
an additional global factor of two-thirds appears in the two-point correlation function
$C^{\mathrm{bb}}(t) = \langle\hat{\mathbf{E}}_i(t)\hat{\mathbf{E}}_j(0)\rangle$, giving
\begin{align}
C^{\mathrm{bb}}(t)
&=\frac{1}{2}\int\limits_{-\infty}^\infty  \mathrm{d} \omega \omega^2 J^{\mathrm{bb}}(\omega)
\left[ C^{\mathrm{bb}}_+(\omega) \mathrm{e}^{\mathrm{i} \omega t}  +
C^{\mathrm{bb}}_-(\omega) \mathrm{e}^{-\mathrm{i} \omega t}\right] \Theta(\omega)
\nonumber \\
&=\frac{1}{2} \int\limits_{-\infty}^\infty  \mathrm{d} \omega \omega^2 J^{\mathrm{bb}}(\omega)
C^{\mathrm{bb}}_+(\omega) \mathrm{e}^{\mathrm{i} \omega t},
\end{align}
where $\beta = 1/ k_{\mathrm{B}}T$,
\begin{equation}
\label{equ:Jblkbdrdtn}
\omega^2 J^{\mathrm{bb}}(\omega) = \frac{2}{3} \hbar \omega^3/ (4\epsilon_0\pi^2c^3),
\end{equation}
$C^{\mathrm{bb}}_+(\omega) = \coth\left(\frac{1}{2} \hbar \omega \beta \right) - 1$,  and
$C^{\mathrm{bb}}_-(\omega) = \left[\coth\left(\frac{1}{2} \hbar \omega \beta \right) +1\right]$.

In obtaining the last expression, the relations $\omega^2 J^{\mathrm{bb}}(-\omega)
= -\omega^2 J^{\mathrm{bb}}(\omega)$ and $C^{\mathrm{bb}}_+(-\omega) = -C^{\mathrm{bb}}_-(\omega)$
have been used.	
Note the cubic dependence of the spectral density on $\omega$, which defines its super-Ohmic
character and which, as  shown below,  allows for small decoherence rates that  support long-lived
coherence.
Indeed, it is responsible for a previously noted but essentially unexplained \cite{TB14},  strong
dependence of the decoherence rate on the system level spacing.

In the long-time regime, $t\rightarrow \infty$,
\begin{equation}
\Gamma^{\mathrm{bb}}_{ab,cd} = \frac{1}{2}\mathsf{M}^{\mathrm{bb}}_{ab,cd}(\omega_{dc})
-
\frac{ \mathrm{i} }{2} \dashint \mathrm{d}\omega\frac{\omega^2 J^{\mathrm{bb}}(\omega)
C^{\mathrm{bb}}_+(\omega)}{\omega-\omega_{dc}}.
\end{equation}
In what follows, the imaginary part of $\Gamma^{\mathrm{bb}}_{ab,cd}(t)$ (spontaneous emission)
is neglected because its contribution, for blackbody radiation, is very small compared
to the real part.

The elements of the Redfield tensor $R^{\mathrm{tb,bb}}_{ab,cd}$ with $a=c$ and $b=c$ are
associated with population transfer, while those with $a\neq b$, $a=c$ and $d=b$ relate to  coherence
dephasing.
For blackbody radiation, these terms read
\begin{align}
\label{equ:Rbb_aabb}
R^{\mathrm{bb}}_{aa,bb} &= 2  \delta_{ab} \sum_e \Gamma^{\mathrm{bb}}_{ae,ea}(\omega_{be})
                      -2 \Gamma^{\mathrm{bb}}_{ba,ab}(\omega_{ba}),
\\
\label{equ:Rbb_abab}
\begin{split}
R^{\mathrm{bb}}_{ab,ab} &= \sum_e \Gamma^{\mathrm{bb}}_{be,ed}(\omega_{be})
                     + \sum_e \Gamma^{\mathrm{bb}}_{ae,eb}(\omega_{ae})
                      - \Gamma^{\mathrm{bb}}_{aa,bb}(0)
                      - \Gamma^{\mathrm{bb}}_{bb,aa}(0),
\\
                     &=  \sum_e \left[ \Gamma^{\mathrm{bb}}_{be,eb}(\omega_{be})
                     + \Gamma^{\mathrm{bb}}_{ae,ea}(\omega_{ae})\right],
\end{split}
\end{align}
Note that $\Gamma^{\mathrm{bb}}_{aa,bb}(0) $ and $\Gamma^{\mathrm{bb}}_{bb,aa}(0)$
vanish in Eq.~(\ref{equ:Rbb_abab}), a result of the fact that the electronic system observable
that is coupled to the radiation field [see Eq.~(\ref{equ:hamiltonian})] is purely non-diagonal
in the excitonic basis.
The vanishing of these terms implies, from an  open quantum systems perspective, that
blackbody radiation does not induce pure dephasing dynamics, i.e.,  $T_2^*=0$.
Hence,  a significant feature of blackbody radiation is that it is population relaxation
(due to stimulated emission) that is the source of coherence dephasing, and is typically slower
than pure dephasing.
%

\subsection{Tensor elements of the quantum dynamical map}
In solving the dynamics in Eq.~(\ref{equ:RFnMME}), the master equation equation is conveniently
written as
\begin{equation}
\label{equ:cmpctME}
\dot{\boldsymbol{\varrho}} = \bar{\mathsf{R}} \boldsymbol{\varrho},
\end{equation}
with elements of the $N^4$-th dimensional vector $\boldsymbol{\varrho}$ given by
$\varrho_{(a-1)N+b} = \rho_{ab}$ and the elements of the $N^4\times N^4$-th dimensional
array
$\bar{\mathsf{R}}$ by $ \bar{R}_{(a-1)N+b,(c-1)N+d} = \mathrm{i}\delta_{ac}\delta_{bd}\omega_{ab}-R_{ab,cd}$.
Since, for the sudden turn-on case, the elements of the Redfield tensor $\mathsf{\bar R}$ are time independent,
the differential matrix equation (\ref{equ:cmpctME}) can be solved exactly, provided that the
eigenvalues and eigenvectors of $\bar{\mathsf{R}}$, the ``damping basis'', can be calculated.
Specifically, if the matrix $\mathsf{A}$ contains the eigenvectors of $\bar{\mathsf{R}}$ and $\mathsf{L}$
its eigenvalues,
$\boldsymbol{\varrho}(t) = \mathsf{A}^{-1}\mathrm{e}^{\mathsf{L} t} \mathsf{A} \boldsymbol{\varrho}(0)$.
Therefore, the eigenvalues of $\bar{\mathsf{R}}$ dictate the time scales of the dynamics, and
the time evolution of each component of $\boldsymbol{\varrho}(t)$ is a superposition of these time
scales.
A similar approach can be carried out for the case of time-dependent $\bar{\mathsf{R}}$-elemets

To allow for effects on any initial state $\boldsymbol{\rho}(0)$, it is convenient to focus
on the quantum dynamical map $\mathsf{\chi}$, which can be reconstructed experimentally \cite{PMA15},
rather than on the time evolution of the density operator \cite{PB13,PB13b,PB14,PB13,PMA15}.
The tensor elements $\chi_{ab,cd}$ of the quantum dynamical map are defined as
\begin{equation}
\chi_{ab,cd}(t) = \sum_j^{N^4} A^{-1}_{(a-1)N+b,j} \left(\mathrm{e}^{\mathsf{L} t} \right)_{jj} A_{j,(c-1)+d},
\end{equation}
so that the time evolution of the density matrix elements obeys the mapping
\begin{equation}
\label{equ:rhooft}
\rho_{ab}(t) = \sum_{cd} \chi_{ab,cd}(t) \rho_{cd}(0).
\end{equation}

Note that if the system is initially in an incoherent mixture of eigenstates, such as in thermal
equilibrium, then the initial density matrix in Eq.~(4) is diagonal, and
a non-vanishing tensor element $\chi_{ab,nn}(t)$ will create coherences due to the coupling
to the vibrational bath or to blackbody radiation.
In particular, if excitation takes place from the ground state, $\hat{\rho}(0) = |g\rangle \langle g| $, then
the elements $\chi_{ab,gg}(t)$ coincide with a density element, i.e.,  $\rho_{ab}(t) = \chi_{ab,gg}(t)$.
The physical meaning of all the tensor elements $\chi_{ab,cd}(t)$ is discussed in detail in
Refs.~\cite{PB13,PB13b,PB14,PB13,PMA15}.

An enhanced understanding of
the important case of slow turn-on of the light, which shows significantly diminished coherence
contributions \cite{DTB16b}, can also be obtained using this open system perspective.
Specifically, as shown in Appendix A,  one way to treat the slow turn-on problem is by artificially introducing time dependent dipole moments, where the growth of excited state population is seen to
slow down and the amplitude of the generated coherences diminish dramatically in an adiabatic process.

\section{Photoinduced Dynamics Results}

Three  examples are considered below as specific applications of this  approach:
the case of (i) a single chromophore, (ii)  a dimer, and (iii) a
subunit of the PC645 antenna system\cite{MD&07}.

\subsection{A Single Chromophore}

A single chromophore, considered as a two level system,  has ground state $\ket{g}$
and excited state $\ket{e}$.
To emphasize the difference in spectral densities,  the
influence of the vibrational bath and the influence of blackbody radiation are considered
separately below.

If the system is only coupled to a vibrational bath of temperature $T^{\textrm{tb}}$ then,
using the Ohmic spectral density
[Eq.~(\ref{equ:spcfcJ})], the eigenvalues $\mathsf{L}^{\mathrm{tb}}$ are given by
\begin{equation}
\label{equ:Ltb}
\mathsf{L}^{\mathrm{tb}} = \{0, 0,
-\mathrm{i} \omega_{ee} - \gamma^{\mathrm{tb}}_{ee},
\mathrm{i} \omega_{ee} - \gamma^{\mathrm{tb}}_{ee} \}
\end{equation}
with $\omega_{eg} = \epsilon/\hbar$ and
\begin{align}
\label{equ:gammatb}
\gamma^{\mathrm{tb}}_{ee} &= \Gamma_{ee,ee}(0)
= 4 k_{\mathrm{B}} T^{\mathrm{tb}} \frac{\tau \Lambda}{\hbar^2}
= 4 \frac{1}{\tau^{\mathrm{tb}}}
\frac{\tau \Lambda}{\hbar}
\end{align}
Here $\tau^{\mathrm{tb}} = \hbar/k_{\mathrm{B}} T^{\mathrm{tb}}$ is the thermal coherence time
of the vibrational bath.
For  convenience below note that the decay component associated with the eigenvalues of
$\mathsf{A}^{\mathrm{tb}}$ is independent of the Rabi frequency $\omega_{eg}$.
The $\mathsf{\chi}$ tensor elements are given by
\begin{equation}
\label{equ:Jtbnmkl}
\begin{split}
\chi_{ee,ee} &= 1, \quad \chi_{ee,eg} = 0, \quad \chi_{ee,ge} = 0, \quad \chi_{ee,gg} = 0,
\\
\chi_{eg,eg} &= \mathrm{e}^{L_4^{\mathrm{tb}}t}, \quad
\chi_{eg,eg} = 0, \quad \chi_{eg,gg} = 0, \quad \chi_{eg,gg} = 0,
\\
\chi_{ge,ee} &= 0, \quad J_{ge,eg} = 0, \quad
\chi_{ge,ge} = \mathrm{e}^{L_3^{\mathrm{tb}}t}, \quad
\chi_{ge,gg} = 0,
\\
\chi_{gg,ee} &= 0, \quad v_{gg,eg} = 0, \quad \chi_{gg,ge} = 0, \quad \chi_{gg,gg} = 1.
\end{split}
\end{equation}
From Eq.~(\ref{equ:rhooft}) it is clear that if  the total population is initially in the ground state,
then the generation of coherences, i.e., non-vanishing $\rho_{eg}(t)$ or $\rho_{ge}(t)$,
would occur via the tensor elements $\chi_{eg,gg}$ and $ \chi_{ge,gg}$.
However, in this case, these are  zero.
That is, no system coherences are generated by coupling a single chromophore to it vibrations.
This result is not altered if the dynamics is generated by a time-dependent Redfield tensor, provided
that through tensor elements that are identically zero remain zero under those circumstance
\cite{PB14,PB13,PMA15}.
However, this model is minimal, with only one excited state with  no coherence allowed between
the ground and (single) excited state.

If only coupling between the system and blackbody radiation is considered, the eigenvalues
$\mathsf{L}^{\mathrm{bb}}$ are given by
\begin{equation}
\label{equ:Lbb}
\mathsf{L}^{\mathrm{bb}} = \{0, -2\gamma^{\mathrm{bb}}_{eg},
-\gamma^{\mathrm{bb}}_{eg} - \mathrm{i}\sqrt{\omega_{eg}^2 - [\gamma^{\mathrm{bb}}_{eg}]^2},
-\gamma^{\mathrm{bb}}_{eg} + \mathrm{i} \sqrt{\omega_{eg} ^2- [\gamma^{\mathrm{bb}}_{eg}]^2} \}
\end{equation}
for $\omega_{eg}^2 - [\gamma^{\mathrm{bb}}_{eg}]^2 >0$ with
\begin{align}
\label{Ref21}
\gamma^{\mathrm{bb}}_{eg} &= R^{\mathrm{bb}}_{eg,eg}
=\Gamma^{\mathrm{bb}}_{eg,ge}(\omega_{eg}) +\Gamma^{\mathrm{bb}}_{ge,eg}(\omega_{ge})
\\ &=
\frac{\boldsymbol{\mu}_{eg}^2} {3 \hbar \pi \epsilon_0 c^3} \omega_{eg}^3
\coth\left(\frac{1}{2} \frac{\hbar \omega_{eg}}{ k_{\mathrm{B}}T^{\mathrm{bb}}} \right) \nonumber
\\
&\sim \frac{k_{\mathrm{B}}T^{\mathrm{bb}}  \boldsymbol{\mu}_{eg}^2}
{3 \hbar^2 \pi \epsilon_0 c^3} \omega_{eg}^2
=\frac{1}{\tau^{\mathrm{bb}}} \frac{ \boldsymbol{\mu}_{eg}^2}
{3 \hbar  \pi \epsilon_0 c^3} \omega_{eg}^2,
\end{align}
with $\boldsymbol{\mu}_{eg} = \langle e |\hat{\boldsymbol{\mu}}|g\rangle$.  Here
 $\tau^{\mathrm{bb}} = \hbar/k_{\mathrm{B}} T^{\mathrm{bb}}$ is the thermal coherence time
of the blackbody radiation.
Note that the approximative expression in Eq. (\ref{Ref21}) assumes a high temperature regime
which simplifies the coth term.
Note also that the eigenvalues in Eq.~(\ref{equ:Lbb}) display a coherent regime
with $(\omega_{eg}^2 - [\gamma^{\mathrm{bb}}_{eg}]^2 )>0$, where oscillations are observed and an incoherent regime with
$(\omega_{eg}^2 - [\gamma^{\mathrm{bb}}_{eg}]^2) <0$, where they are not. There is
no analogous transition for a system coupled to a vibrational bath.

The cubic dependence (square dependence at high temperature) on
$\omega_{eg}$ of $\gamma^{\mathrm{bb}}_{eg}$  in Eq. (\ref{Ref21})        is a manifestation of the
super-Ohmic character of blackbody radiation [see Eq.~(\ref{equ:Jblkbdrdtn})].
Interestingly, for small energy gaps $\hbar \omega_{eg}$, the decoherence rate may be substantially
smaller than that of the  Ohmic spectral density in Eq.~(\ref{equ:spcfcJ}) for which the
decoherence rate is $\omega_{eg}$-independent [see Eq.~(\ref{equ:gammatb})].
Indeed, this characteristic feature of blackbody radiation is often neglected
 and the radiation  is  described  as  white noise with a constant dephasing rate.   However, this $\omega_{eg}$ dependence
may be significant in the case of
hyperfine-transitions-based atomic clocks  (see, e.g., \cite{JH&14}) and, as shown below,
it is  important in the description of excitonic energy transfer.

The tensor elements that describe the interaction of the system  with blackbody radiation are
\begin{equation}
\label{equ:Jbbnmkl}
\begin{split}
\chi_{ee,ee}^{\mathrm{bb}} &= \frac{1}{2}(1+\mathrm{e}^{L_2^{\mathrm{bb}}t}), \quad
\chi_{ee,eg}^{\mathrm{bb}} = 0, \quad
\chi_{ee,ge}^{\mathrm{bb}} = 0, \quad
\chi_{ee,gg}^{\mathrm{bb}} = \frac{1}{2}(1-\mathrm{e}^{L_2^{\mathrm{bb}}t}),
\\
\chi_{eg,ee}^{\mathrm{bb}} &= 0, \quad
\chi_{eg,eg}^{\mathrm{bb}} = \frac{1}{2}(1-\zeta_{eg}) \mathrm{e}^{L_3^{\mathrm{bb}}t} +
\frac{1}{2}(1+\zeta_{eg}) \mathrm{e}^{L_4^{\mathrm{bb}}t}, \quad
\chi_{eg,ge}^{\mathrm{bb}} = -\chi^{\mathrm{bb}}_{ge,eg}, \quad
\chi_{eg,gg}^{\mathrm{bb}} = 0,
\\
\chi_{ge,ee}^{\mathrm{bb}} &= 0, \quad
\chi_{ge,eg}^{\mathrm{bb}} = \frac{1}{2\mathrm{i}} \frac{\gamma^{\mathrm{bb}}_{eg}}{\omega_{eg}}
(\mathrm{e}^{L_4^{\mathrm{bb}}t} - \mathrm{e}^{L_3^{\mathrm{bb}}t}), \quad
\chi_{ge,ge}^{\mathrm{bb}} = \chi^{\mathrm{bb}*}_{eg,eg} , \quad
\chi_{ge,gg}^{\mathrm{bb}} = \frac{1}{2}(1-\mathrm{e}^{L_2^{\mathrm{bb}}t}),
\\
\chi_{gg,ee}^{\mathrm{bb}} &= 0, \quad
\chi_{gg,eg}^{\mathrm{bb}} = 0, \quad
\chi_{gg,ge}^{\mathrm{bb}} = 0, \quad
\chi_{gg,gg}^{\mathrm{bb}} = \frac{1}{2}(1+\mathrm{e}^{L_2^{\mathrm{tb}}t}).
\end{split}
\end{equation}
with $\zeta_{eg} = \varpi_{eg}/\omega_{eg}$ and
$\varpi_{eg} = \sqrt{\omega_{eg}^2 - [\gamma^{\mathrm{bb}}_{eg}]^2}$.
From $\mathsf{L}^{\mathrm{bb}}$ and $\mathsf{\chi}^{\mathrm{bb}}$ is clear that, as
anticipated above, the standard free-pure-dephasing relation holds for the population
decay-rate $k^{\mathrm{bb}}_{eg}$ and the decoherence rate $\gamma^{\mathrm{bb}}_{eg}$,
namely $\gamma^{\mathrm{bb}}_{eg} = R^{\mathrm{bb}}_{eg,eg}=
\frac{1}{2}k^{\mathrm{bb}}_{eg}$.  The fact that the decoherence rate is one-half the
stimulated emission rate emphasizes the unusually slow rate of decoherence due to
blackbody radiation.

\subsection{Model PC645}

As a second model system, consider a subunit of the protein-antenna phycocyanin
PC645 of marine cryptophyte algae, where long-lasting coherence in 2DPE experiments
examining energy transfer from DBV to MBV chromophores,  has been reported \cite{CW&10}.
Our earlier results   \cite{JB91,MV10,PB11,TB14,DTB16,DTB16b}
have made clear that such coherences
are a consequence of the use of pulsed laser excitation, and that the primary information gained
was information on the system-bath interaction post excitation. Here we examine the more
realistic case where, after rapid excitation, the system continues to interact with the
incident blackbody radiation.
\begin{table}[h]
\begin{tabular}{c c c}
\hline
\multirow{2}{*}{Chromophore}  & Transition & Transition dipole
\\
& energy [eV] &  moment [D]
\\
\hline
DBV 50/61 C &  2.112 & 13.2
\\
DBV 50/61 D &  2.122 & 13.1
\\
MBV 19 A &  1.990 & 14.5
\\
MBV 19 B & 2.030 & 14.5
\\
\hline
\end{tabular}
\caption{Parameters for the chromophores in PC645 \cite{MD&07,HW&14}}
\label{tab:paramPC645}
\end{table}
Specifically, consider the two dihydrobiliverdin (DBV) chromophores
and the two meso-biliverdin (MBV) chromophores.
The magnitude and orientation of the transition dipole moments, for this particular antenna
system, were recently calculated \cite{HW&14}.
For the purpose of the present discussion, it suffices to consider them parallel, and with
magnitude as given experimentally in Ref.~\cite{MD&07}.
Table \ref{tab:paramPC645} summarizes this information.  We examine several cases.

\subsection{A Dimer System}

Before proceeding with the four chromophore example, it is illustrative
 to  consider the dynamics induced in a dimer by incoherent light,
specifically, the DBV dimer of PC645  (see Table~\ref{tab:paramPC645}).
This system has a ground state $\ket{g}$, two single exciton states $\{\ket{e},\ket{e^\prime}\}$
and a two-exciton state $\ket{f}$. The energy gap between the donor and acceptor
excited state energy levels is denoted $\Delta$.
%

%
%

To focus only on the action of the radiation, the decoherence effects of the
vibrations on the excited states are set to zero in Eq.~(\ref{equ:hamiltonian}).
For example, for excitation from the ground state,
$\hat{\rho}(0)=|g\rangle \langle g|$, $\hat{\rho}_{ab}(t) = \chi^{\mathrm{tb,bb}}_{ab,gg}(t)$.
Hence, the $\chi^{\mathrm{tb,bb}}$ matrix element satisfies a relation analogous to
Eq.~(\ref{equ:RFnMME}), i.e.,
$\dot{\chi}^{\mathrm{tb,bb}}_{ab,gg} = \sum_{cd} R^{\mathrm{tb,bb}}_{ab,cd}\chi^{\mathrm{tb,bb}}_{cd,gg}$.
Furthermore,
following the procedure introduced in the derivation of Eq.~(\ref{equ:cmpctME}), it is possible
to obtain an  expression for $\chi^{\mathrm{tb,bb}}_{ab,cd}$ for the case of excitation from an arbitrary state.

\begin{figure*}[h]
\includegraphics[width=0.49\columnwidth]{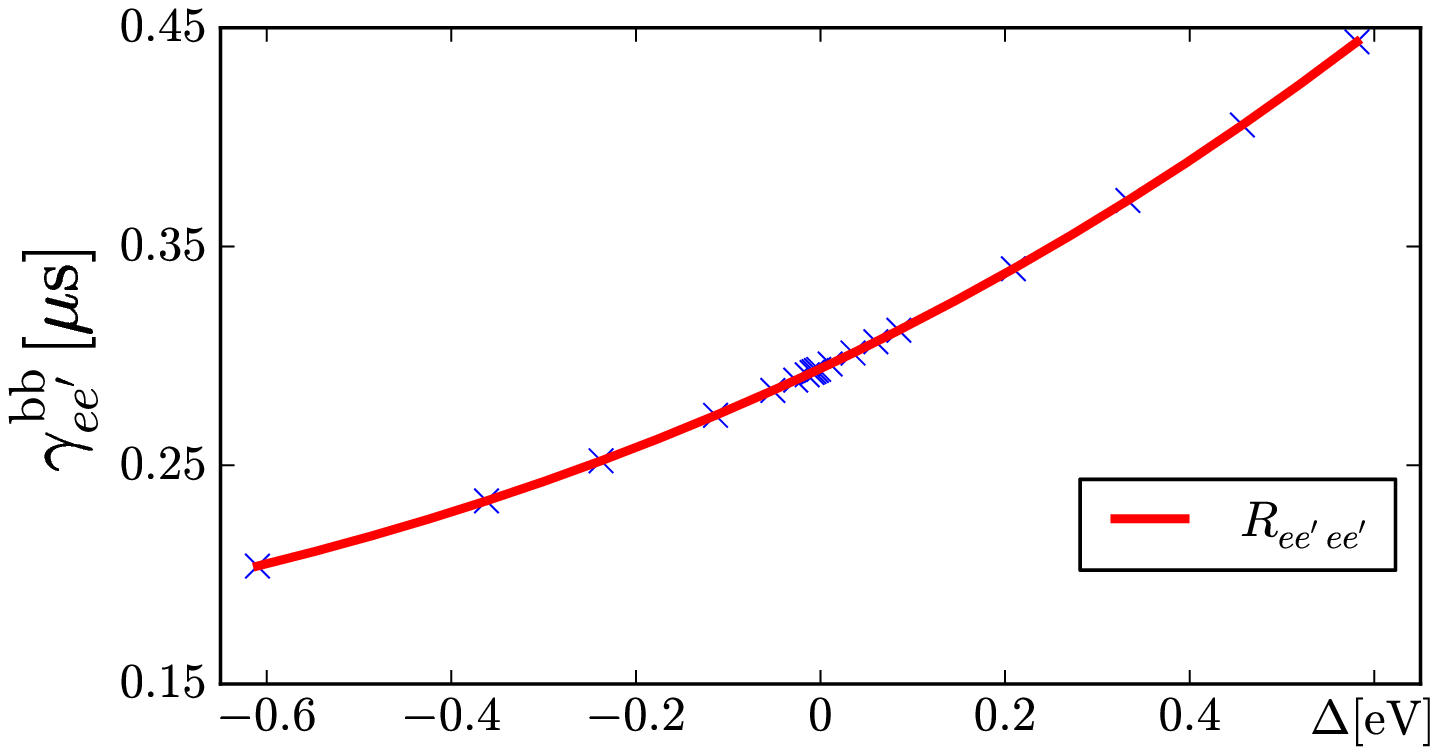}
\includegraphics[width=0.49\columnwidth]{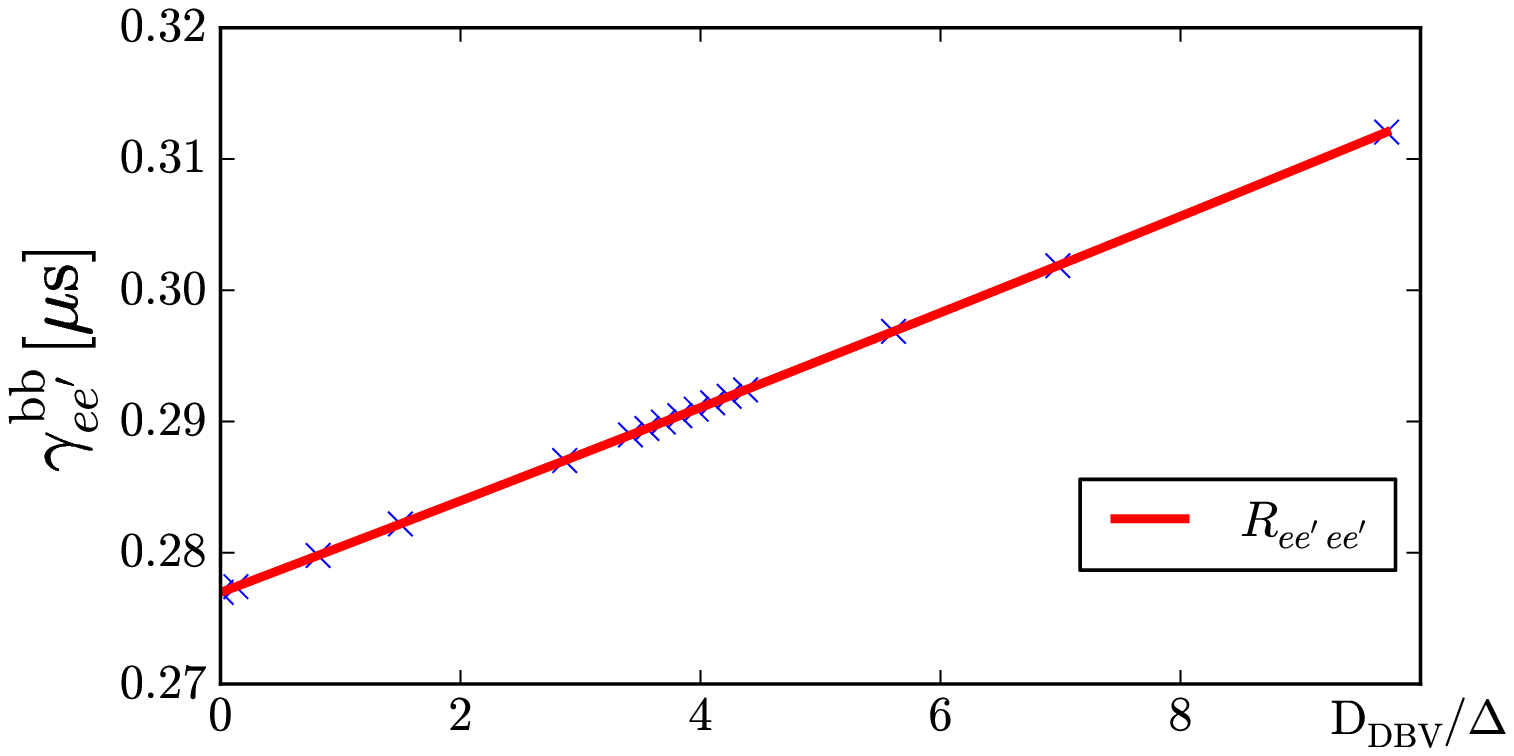}
\caption{
Left panel: $\gamma^{\mathrm{bb}}_{ee^\prime}$ as a function of the donor-acceptor 
excited state energy gap $\Delta$
(blue marks) and $R^{\mathrm{bb}}_{e e^\prime e e^\prime}$
as a function of  $\Delta$ (red curve).
Right panel: $\gamma^{\mathrm{bb}}_{ee^\prime}$ as a function of the ratio between the
coupling strength $\textrm{D}_{\mathrm{DBV}}$ of the DBV-dimer and the energy gap $\Delta$
(blue marks) and $R^{\mathrm{bb}}_{e e^\prime e e^\prime}$ as a function of the ratio
$\textrm{D}_{12}/\Delta$ (red curve).
Other parameters are set to simulate the DBV dimer in Table~\ref{tab:paramPC645}
and with $T_{\mathrm{bb}} = 5600$~K.
}
\label{fig:gammaeep}
\end{figure*}

Interest here is the decay time of the coherence established in the single-exciton manifold
$\rho_{ee^\prime}(t)$ for excitation from the ground state with blackbody radiation.
The main contribution to  $\chi^{\mathrm{bb}}_{ee^\prime,gg}$ results from
the component with the eigenvalue $L^{\mathrm{bb}}_{ee^\prime}$ of
$\bar{\mathsf{R}}^{\mathrm{bb}}$ that leads to oscillations at the frequency
$\omega_{ee^\prime} \sim \Im \lambda^{\mathrm{bb}}_{ee^\prime}$ with decay rate
$\gamma^{\mathrm{bb}}_{ee^\prime} = |\Re \lambda^{\mathrm{bb}}_{ee^\prime}|$.
Figure~\ref{fig:gammaeep} depicts the functional dependence of $\gamma^{\mathrm{bb}}_{ee^\prime}$
on the energy gap $\Delta$ (left hand side panel) and
on the scaled coupling  constant $\textrm{D}_{12}/\Delta$ (right-hand panel)  for typical parameters
in light-harvesting systems (see Table~\ref{tab:paramPC645} above).

\begin{figure*}[h]
\includegraphics[width=0.49\columnwidth]{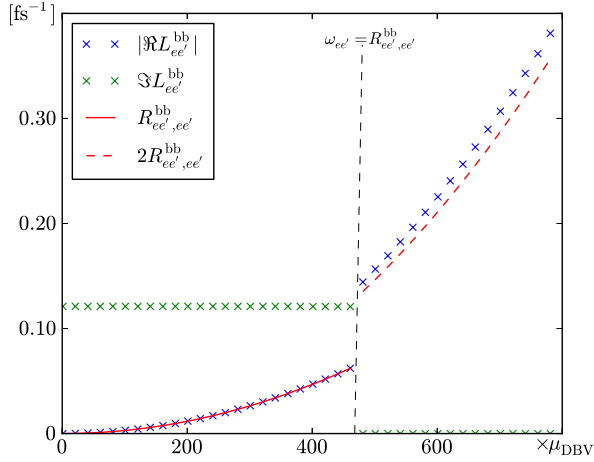}
\caption{
Real (blue marks) and imaginary (green marks) parts of eigenvalue $L_{ee^\prime}$
as a function of the transition dipole moment $\mu_{\mathrm{DBV}}$
of the DBV-dimer.
The continuous red curve depicts $R^{\mathrm{bb}}_{e e^\prime e e^\prime}$ as a
function of $\mu_{\mathrm{DBV}}$ whereas the dashed red curve does so for
$2R^{\mathrm{bb}}_{e e^\prime, e e^\prime}$.
Other parameters are set to simulate the DBV dimer in Table~\ref{tab:paramPC645}
and with $T_{\mathrm{bb}} = 5600$~K.
}
\label{fig:trnstin}
\end{figure*}

Further insight is afforded by Fig.~\ref{fig:trnstin}. For the present case and at arbitrary temperature it is
possible to identify  two well-defined dynamical regimes: (i) A coherent regime for
$\omega_{ee^\prime} > R^{\mathrm{bb}}_{ee^\prime,ee^\prime}$ that guaranties that
$L_{ee^\prime}$ has non-vanishing imaginary part and
(ii) an incoherent regime for $\omega_{ee\prime} < R^{\mathrm{bb}}_{ee^\prime,ee^\prime}$
that defines a purely real eigenvalue $L_{ee^\prime}$.
The onset of the transition as a function of $R^{\mathrm{bb}}_{ee^\prime,ee^\prime}$ coincides
with that found for the case of a single chromophore above, as discussed in Eq. (\ref{Ref21}),
These estimates provide additional insight into the ``underdamped" and ``overdamped"
regions noted previously \cite{DTB16,DTB16b}.
From Figs.~\ref{fig:gammaeep} and \ref{fig:trnstin}, it is clear that $R^{\mathrm{bb}}_{e e^\prime e e^\prime}$
accounts quantitatively for the decay rate $\gamma_{e e^\prime}^{bb}$
of the superposition between $|e\rangle$ and $|e^\prime\rangle$.
Moreover, in the incoherent regime the transfer rate can be approximated by $2R^{\mathrm{bb}}_{e e^\prime e e^\prime}$.


In photosynthetic light-harvesting complexes (see Table \ref{tab:paramPC645}), energy transfer
occurs between eigenstates that are very close to one another, e.g., $\omega_{ab}\sim 10^2$cm$^{-1}$,
and with transition dipole moments of the order of $10$~D.
Thus, for \emph{isolated} light-harvesting complexes, suddenly turned-on incoherent
light induced dynamics are coherent (see below) provided
that $\gamma^{\mathrm{bb}}_{ab} / \omega_{ab} < 1$. By contrast,
for atomic or molecular transitions with $\omega_{ab}$ on the order of $10^3$ cm$^{-1}$,
with transition dipole moments on the order of $10$~D, incoherent dynamics are expected.

\begin{figure*}[h]
\includegraphics[width=0.49\columnwidth]{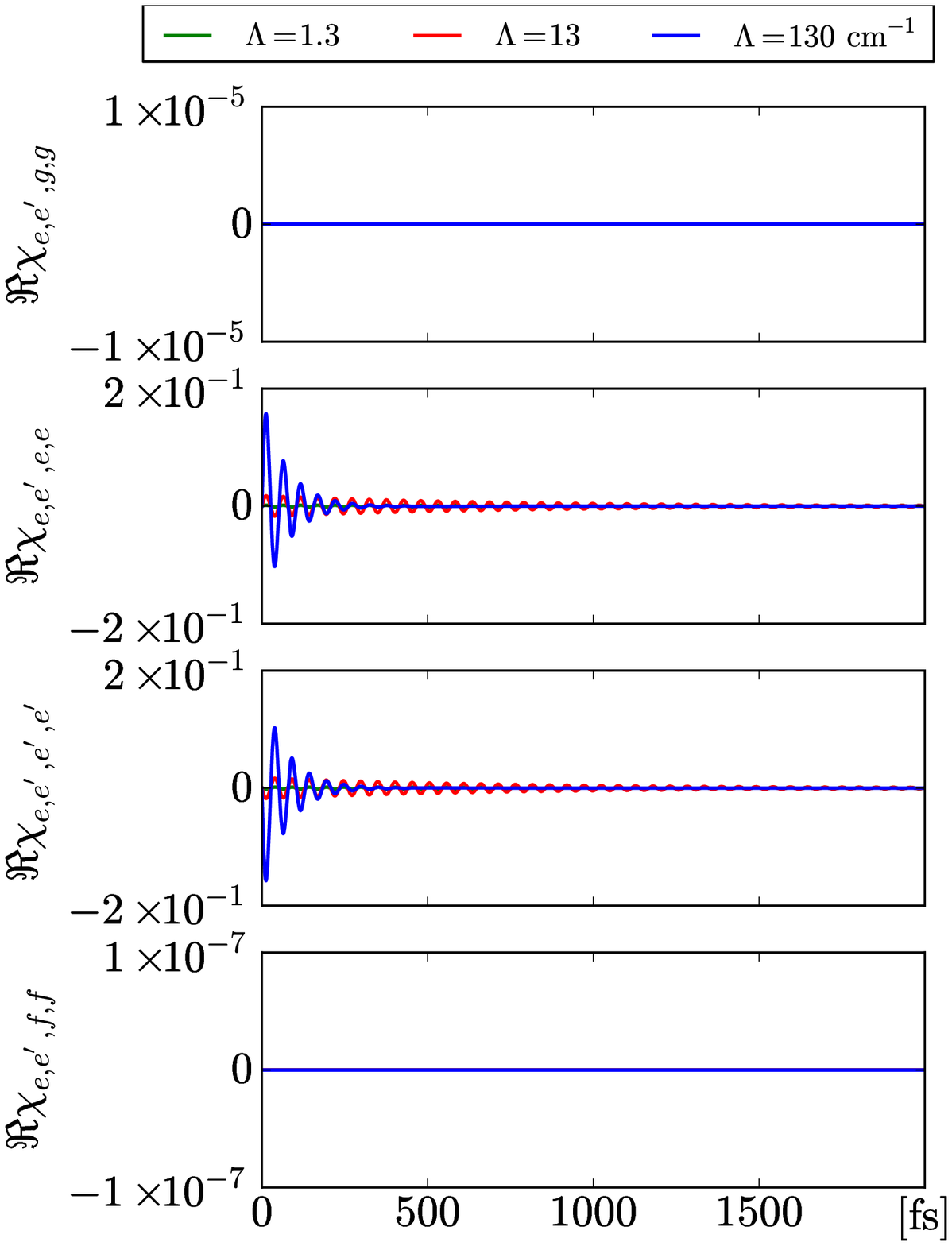}
\includegraphics[width=0.49\columnwidth]{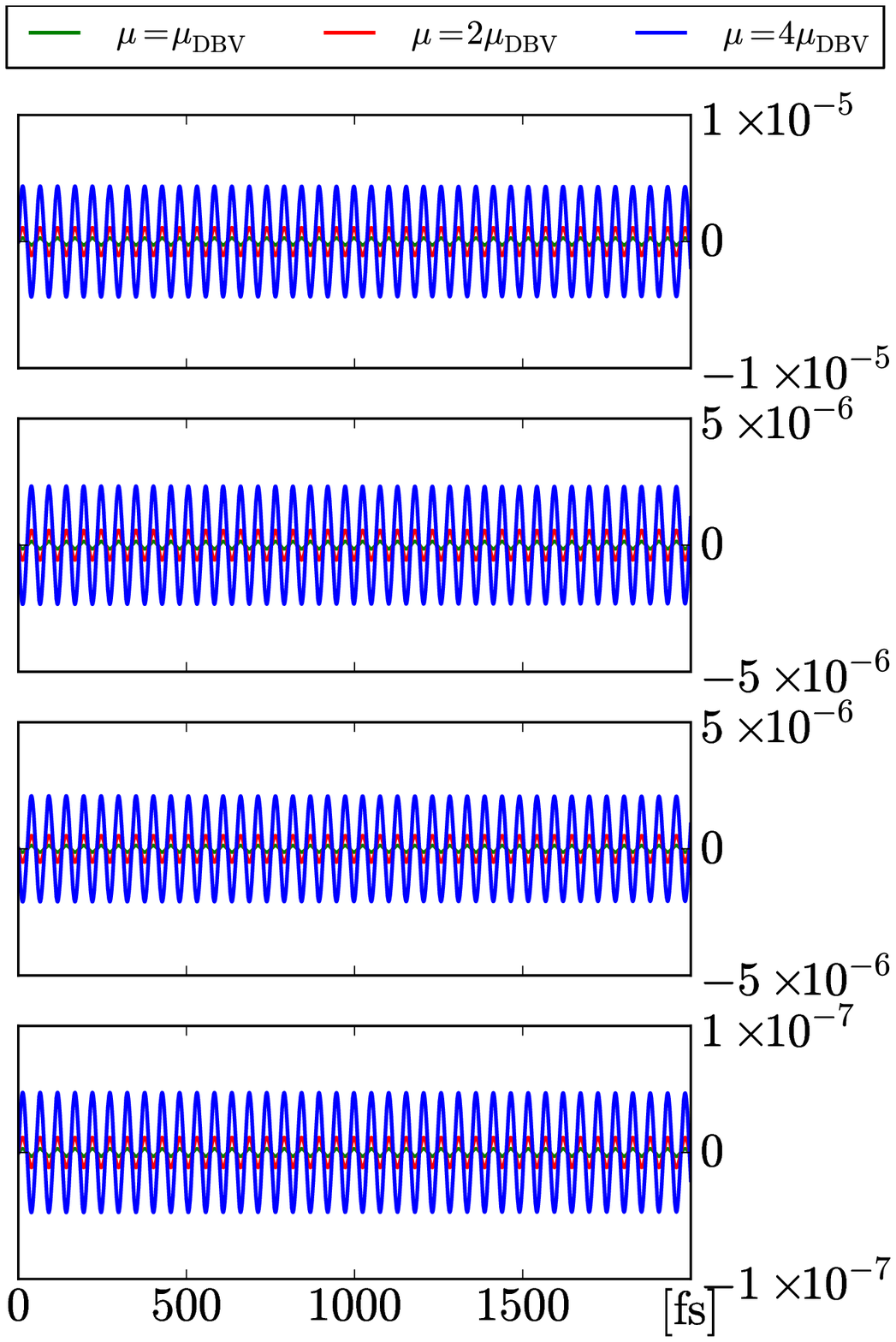}
\vspace{-0.5cm}
\caption{
Left panel: Tensor elements $\chi^{\mathrm{tb}}_{ee^\prime,aa}$
in the absence of blackbody radiation for different values of the reorganization
energy $\Lambda$.
Right panel: Tensor elements $\chi^{\mathrm{bb}}_{ee^\prime,aa}$
in the absence of the thermal bath radiation for different values of the in site dipole
transition moments $\mu$.
Parameters are $T_{\mathrm{tb}} = 300$~K, $T_{\mathrm{bb}} = 5600$~K, and colour
coding for different $\Lambda$ values is shown on top.
}
\label{fig:JnmklTBa}
\end{figure*}

\begin{figure*}[h]
\includegraphics[width=0.49\columnwidth]{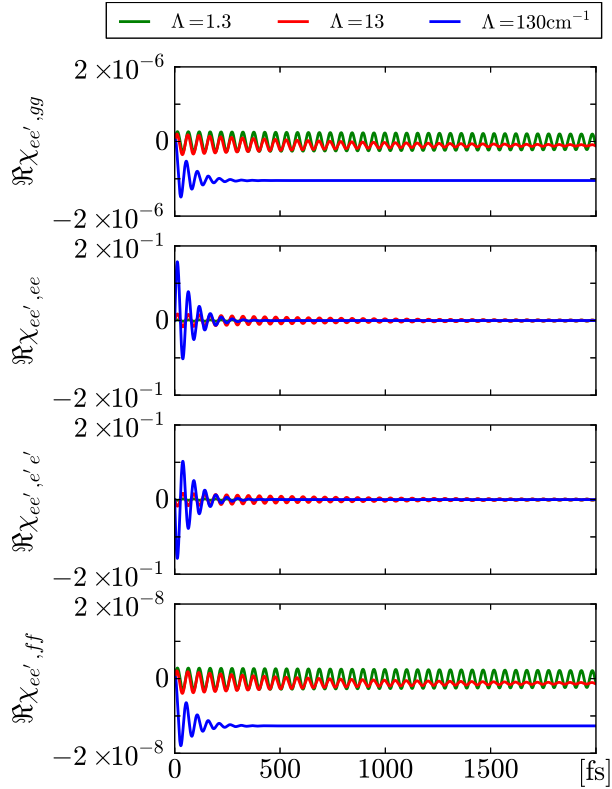}
\vspace{-0.5cm}
\caption{Tensor elements $\chi^{\mathrm{tb+bb}}_{ee^\prime,aa}$
for different values of the reorganization energy $\Lambda$.
%
%
Parameters are $T^{\mathrm{tb}} = 300$~K, $T^{\mathrm{bb}} = 5600$~K, and colour coding for different $\Lambda$ values is shown on top.
}
\label{fig:JnmklTBb}
\end{figure*}

Figures~\ref{fig:JnmklTBa} and
\ref{fig:JnmklTBb} display the tensor elements $\chi_{ee^\prime,gg}$,
$\chi_{ee^\prime,ee}$, $\chi_{ee^\prime,e^\prime e^\prime}$, $\chi_{ee^\prime,ff}$
for various cases associated with generating coherences in the singly excited manifold
$\{e,e^\prime\}$ from incoherent initial states.
Specifically, the left panel of Fig.~\ref{fig:JnmklTBa} shows the results in the
absence of blackbody radiation, i.e., the tensor elements $\chi^{\mathrm{tb}}_{ee^\prime,aa}$,
whereas the right panel displays the results in the absence of the thermal bath,
i.e., the tensor elements $\chi^{\mathrm{bb}}_{ee^\prime,aa}$.
A number of significant features, and differences between blackbody and phonon baths, are evident.
(i) The thermal bath introduced in Eq.~(\ref{equ:hamiltonian}) can not
excite coherences in the singly excited manifold given population initially in the ground
or doubly excited state;
(ii) the tensor elements induced by the thermal bath in the singly excited manifold,
$\chi^{\mathrm{tb}}_{ee^\prime,ee}$ and $\chi^{\mathrm{tb}}_{e^\prime e^\prime,aa}$,
are five order of magnitude larger than those of blackbody radiation;
(iii) the tensor elements induced by the thermal bath in the singly excited manifold,
$\chi^{\mathrm{tb}}_{ee^\prime,ee}$ and $\chi^{\mathrm{tb}}_{e e^\prime, e^\prime e^\prime}$,
decay far faster than those induced by blackbody radiation.
This suggest that excitation with suddenly turned on incoherent light, can be considered
as an effective coherent excitation over timescales less than the thermal bath decoherence
times.
(iv) The tensor elements $\chi^{\mathrm{tb}}_{ee^\prime,ee}$ and
$\chi^{\mathrm{tb}}_{e e^\prime, e^\prime e^\prime}$ are out of phase with one another and the
elements $\chi^{\mathrm{bb}}_{ee^\prime,ee}$ and $\chi^{\mathrm{bb}}_{e e^\prime, e^\prime e^\prime}$
oscillate in phase.
This characteristic may well be relevant for identifying electronic from vibrational
coherences in light-harvesting systems, a crucial problem in 2DPE light-harvesting experiments
\cite{TPJ13}.

The left panel of Fig.~\ref{fig:JnmklTBb} displays the results in the
presence of both blackbody radiation and the thermal bath, i.e., the tensor
elements $\chi^{\mathrm{tb+bb}}_{ee^\prime,aa}$.
The results are relatively insensitive to the dipole transition moments: increasing them
by a factor of four, for example, gave qualitatively similar results. 

The long time appearance of stationary coherences assisted by $\chi^{\mathrm{tb+bb}}_{ee^\prime,gg}$
and $\chi^{\mathrm{tb+bb}}_{ee^\prime,ff}$ is evident in the topmost and lowest panels in Fig.~\ref{fig:JnmklTBb}.
In accord with \cite{PB13,PB13b}, the amplitude of these coherence increase with increasing
coupling to the vibrations.

\subsection{The PC645 antenna complex}
For the multilevel system configuration discussed here [see Eq.~(\ref{equ:hamiltonian})], and for the case of PC645,
the dynamics grows in complexity 
due to the various dipole transitions induced by
the incident light, and the multiple interactions between energy eigenstates excited by sunlight.
For this multilevel configuration, the populations and decoherence rates are function of the
coupling and energy gaps so that it is not straightforward to anticipate the behaviour of these rates.
However, even this case, the approximation $\Re L^{\mathrm{bb}}_{ab}\sim R^{\mathrm{bb}}_{ab,ab}$
holds.
%
%
%


\textit{Excitation under model conditions}.
Consider first the case where all chromophores are initially in the ground state, and where
they are initially decoupled from the environment (the latter is, indeed, an approximation, as discussed
in detail in \cite{PB13}), and where the suddenly turned on excitation by sunlight induces the dynamics.

For the ground state of PC645,  the Hilbert
space is spanned by 16 energy eigenstates $\{|e_1\rangle,|e_2\rangle,... |e_{16}\rangle\}$ with eigenvalues
$e_1 > e_2 > ... > e_{16}$.
The single exciton manifold is spanned by the projectors
$\{ \proj{e_{12}}{e_{12}}, \proj{e_{13}}{e_{13}}, \proj{e_{14}}{e_{14}}, \proj{e_{15}}{e_{15}}\}$
which are delocalized over the sites.
Excitations in chromophores DBVd or DBVc would correspond to populations in   $\proj{e_{12}}{e_{12}}$ and
$\proj{e_{13}}{e_{13}}$, while excitations in the chromophores MBVb and MBVa mainly involve
 $\proj{e_{14}}{e_{14}}$ and $\proj{e_{15}}{e_{15}}$.
\begin{figure*}[h]
\includegraphics[width=0.49\columnwidth]{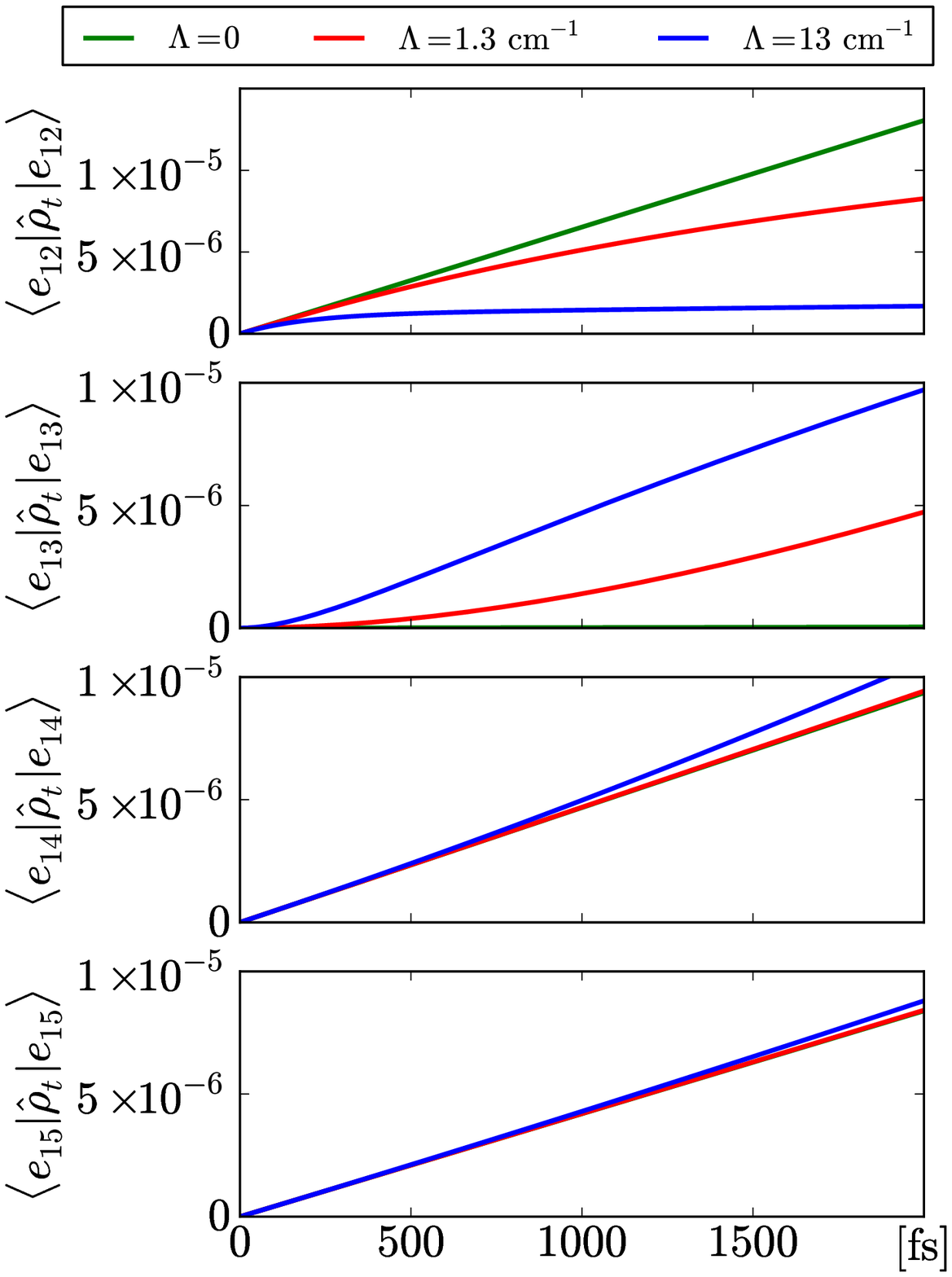}
\includegraphics[width=0.49\columnwidth]{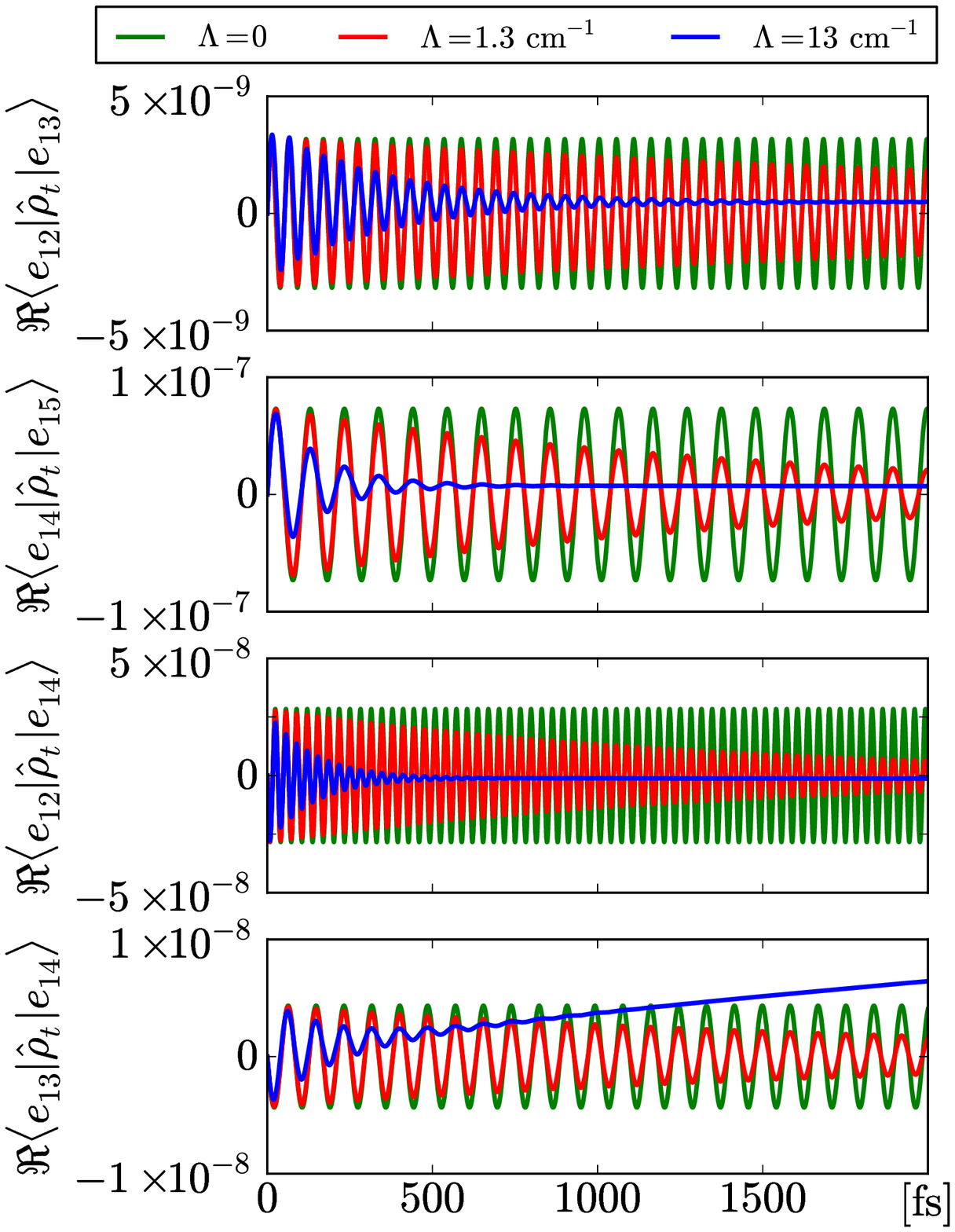}
\vspace{-0.5cm}
\caption{Left
panel: Time evolution at room temperature of the states associated with the single exciton manifold
for different values of $\Lambda$.
Right panel: Time evolution at room temperature of coherences  of sample superpositions between
single-exciton-manifold states induced by suddenly turned on sunlight.
Parameters are $\lambda = 100$~cm$^{-1}$, $T_{\mathrm{env}} = 300$~K, $T_{\mathrm{bb}} = 5600$~K
and $\hat{\rho}(t=0)=\proj{e_{16}}{e_{16}}$, i.e., in the ground state.}
\label{fig:excbasisnatural}
\end{figure*}

Figure \ref{fig:excbasisnatural} shows the time evolution of the populations of the eigenstates
associated with the single exciton manifold as well as some of the coherent superpositions prepared
between these states
by suddenly turned on incoherent light (sunlight).
The observed linear population growth is expected in low intensity incoherent light.
If there is
no coupling to the local environments ($\Lambda = 0$), suddenly turned on sunlight is seen to  prepare coherent
superpositions that last for hundreds of picoseconds.
However, the amplitude of the coherences is approximately two orders of magnitude smaller
than that of the excited state population and so becomes irrelevant quickly\cite{SB14}.
Further, when the coupling to the environment is included ($\Lambda \neq 0$), the coherent
superpositions maintained in sunlight decay due to the underlying incoherent dynamics of
the local environment.
In this case, due to the initial condition (all the population in the ground state), no coherences between the
ground state and the excited states are present, i.e., coherences in this case are among
excited states.

The left panel of Fig.~\ref{fig:excbasisnatural} also displays
an interesting dependence on $\Lambda$. For example, for
$\langle e_{12}|\hat\rho_t|e_{12}\rangle$  absorption  increases with
increasing value of $\Lambda$, whereas the opposite is the case
for  $\langle e_{13}|\hat\rho_t|e_{13}\rangle$.
These  considerable differences in the state populations are a manifestation of the
effect of the coupling of the system to the bath and the
associated flow of population between eigenstates and trace preservation of $\hat{\rho}(t)$.

\begin{figure*}[h]
\includegraphics[width=0.49\columnwidth]{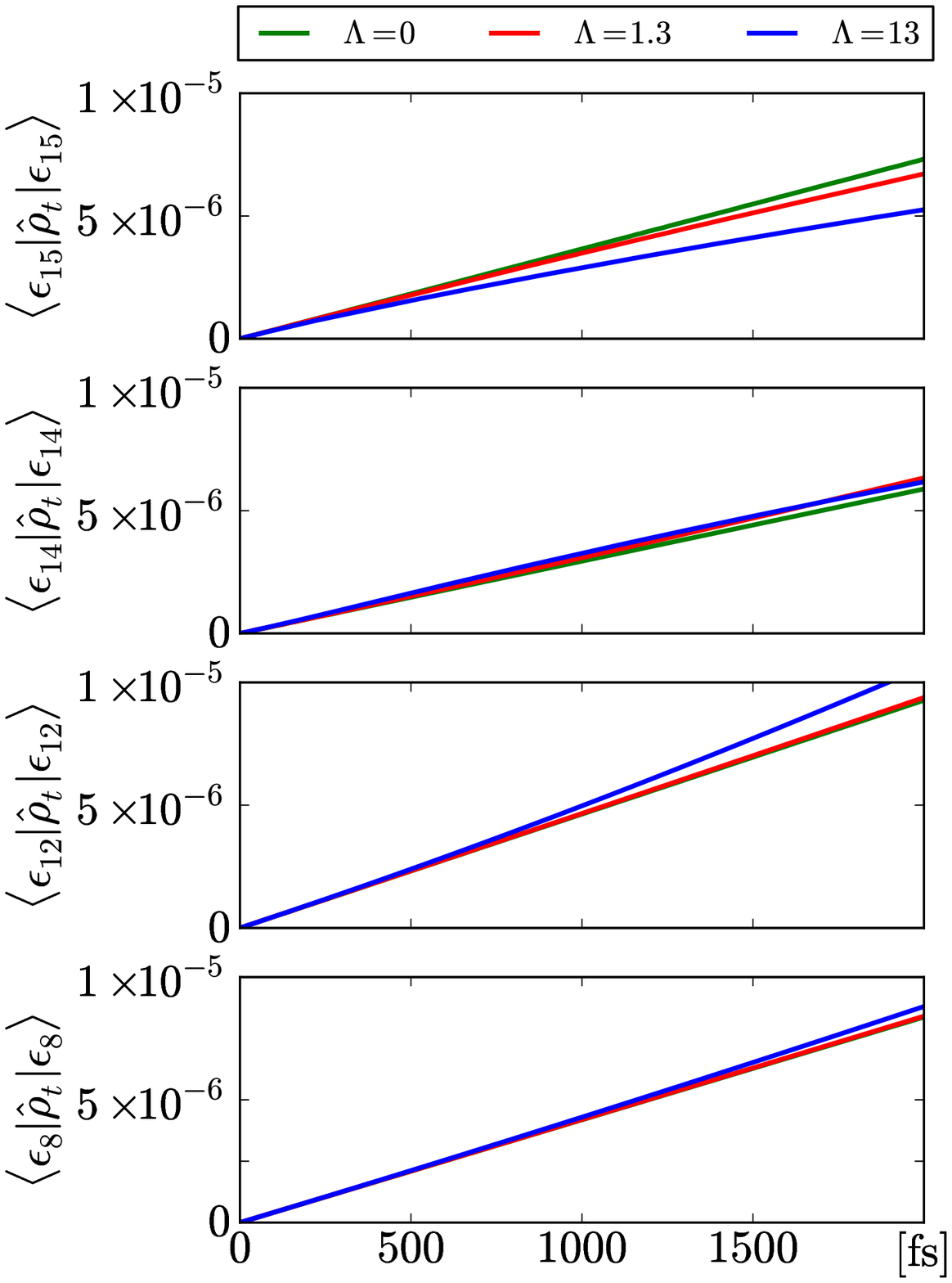}
\includegraphics[width=0.49\columnwidth]{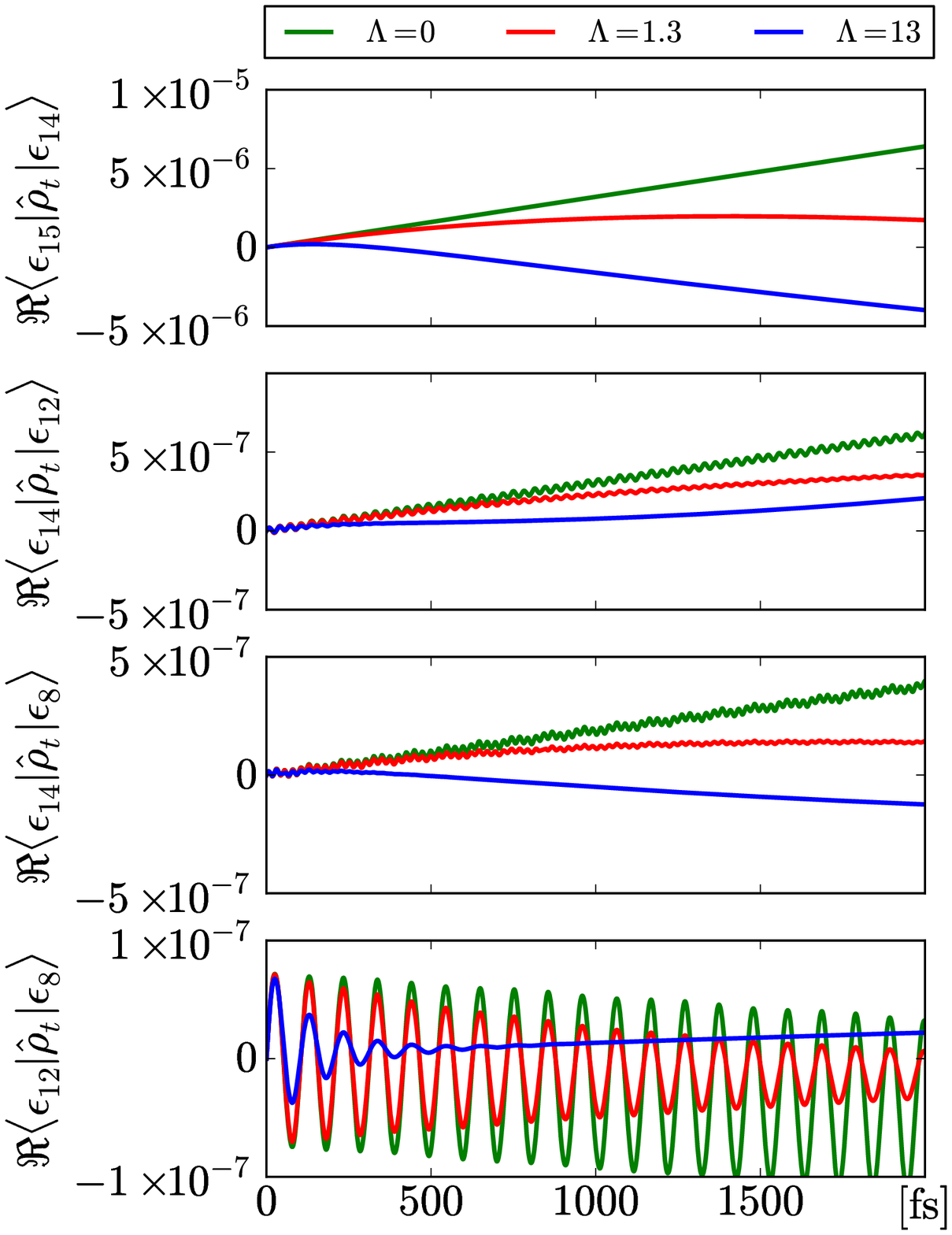}
\vspace{-0.5cm}
\caption{Description and parameters as in Fig.~\ref{fig:excbasisnatural} but with matrix
elements  in the site basis.
Note that the populations of the states
$\{|\epsilon_8\rangle   \langle \epsilon_8|, |\epsilon_{12}\rangle   \langle \epsilon_{12},| \epsilon_{14}
\rangle \langle \epsilon_{14}, |\epsilon_{15}\rangle \langle \epsilon_{15}\}$
refer to  the chromophores MBVa, MBVb, DBVc and DBVd, respectively.
}
\label{fig:sitebasisnatural}
\end{figure*}

Also of interest is the behaviour of populations and coherences
as viewed from the site  basis. To this  end, we have
transformed  from  the  eigenstate  to  the site basis, denoted
$\{|\epsilon_1\rangle,| \epsilon_2\rangle, \ldots\}$.
In this basis the single exciton manifold is spanned by the projectors
$\{ \proj{\epsilon_8}{\epsilon_8}, \proj{\epsilon_{12}}{\epsilon_{12}}$,
$\proj{\epsilon_{14}}{\epsilon_{14}}, \proj{\epsilon_{15}}{\epsilon_{15}}\}$, which correspond to the chromophores
MBVa, MBVb, DBVc and DBVd, respectively.
Figure \ref{fig:sitebasisnatural} shows the time evolution of site-basis matrix elements associated with the
single exciton manifold presented in Fig.~\ref{fig:excbasisnatural}.
In this case, coherences are approximately an order of magnitude larger than in the exciton basis, but
the  populations are of similar magnitude. Hence, the coherences become negligible as time progresses.

One note is in order.
For larger values of the reorganization energy $\Lambda$, stationary coherences in the long-time limit in the  eigenstate
 basis are expected to be larger \cite{PB13}.
These coherences are  a direct  consequence of the
coupling of the electronic degree of freedom to the vibrational bath \cite{PB13}
discussed above.
Since the focus here is on the nature of the excited states prepared by sunlight, larger values of the
reorganization values are not considered.

\textit{Energy pathways under incoherent and coherent radiation.}
As we have previously argued \cite{BS12} dynamics observed in pulsed laser experiments is
distinctly different from that observed in nature due to the {\it coherence characteristics}
of the incident light. Here we note a second significant difference arising from the
{\it incident spectrum} of the light. In particular,
2DPE studies have focused on the spectral region associated with absorption in
the DBV$_-$ state (see the Fig.~1.C in Ref.~\cite{CW&10}), with  energy then flowing  towards 
the MBV molecules \cite{CW&10}. 
However, in absence of vibrations, as shown below, this is not the  only pathway associated 
with natural light absorption, where the excitation has a wider frequency spectrum.

For example, for the ideal unitary case of $\Lambda=0$ in Fig.~\ref{fig:sitebasisnatural},
it is clear that the four
chromophores DBVc, DBVd, MBVa, MBVb show essentially the same rate of population growth,
 all  reaching  $\sim 5\times10^{-6}$ in 1~ps.
\begin{figure*}[h]
\includegraphics[width=0.49\columnwidth]{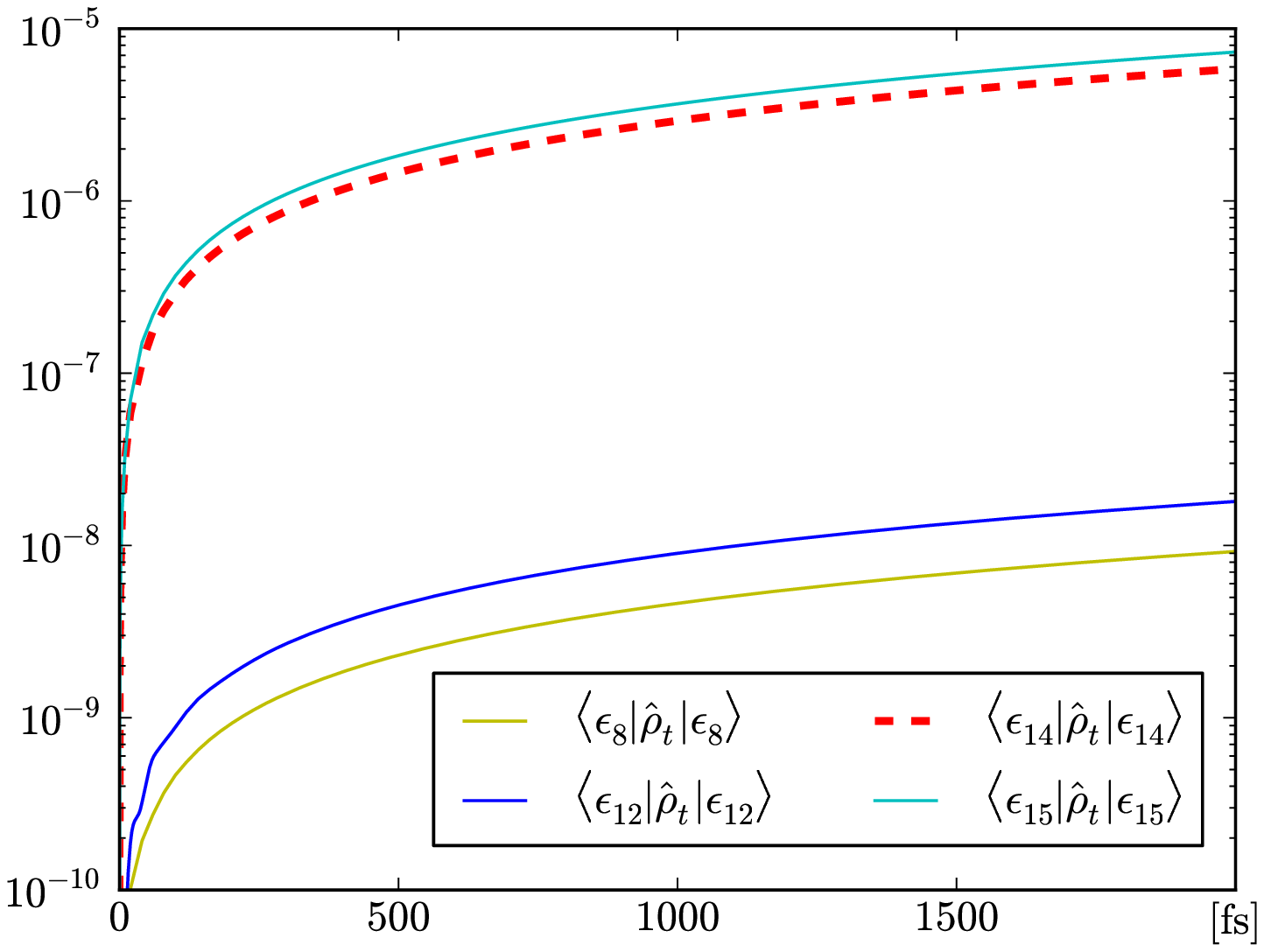}
\includegraphics[width=0.49\columnwidth]{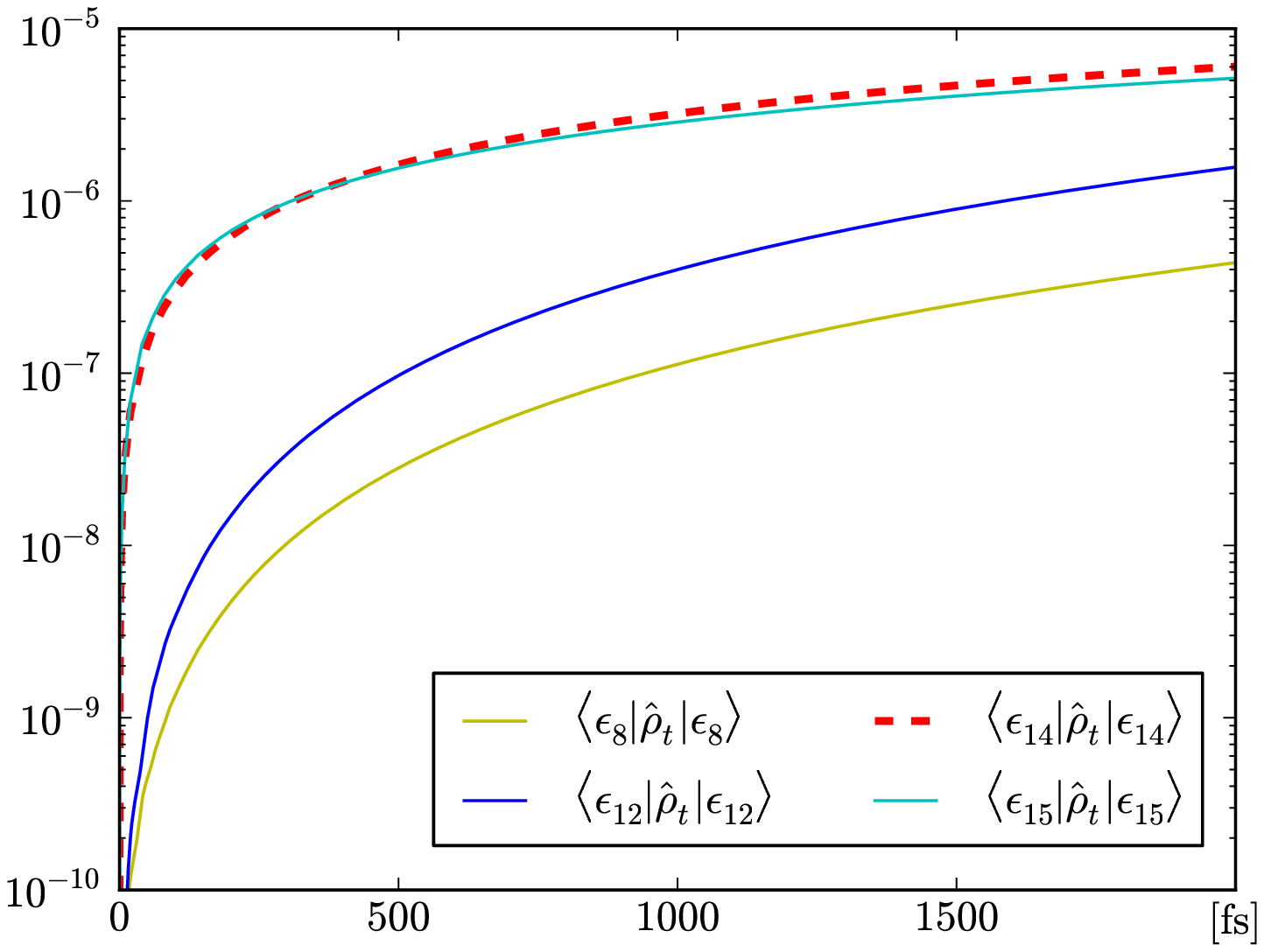}
\vspace{-0.5cm}
\caption{Time evolution the states  associated with the single exciton manifold in the site basis
with no coupling of the MBV molecules to the thermal light and with  $\Lambda=0$ (left panel) 
and $\Lambda=13$~cm$^{-1}$ (right panel).
Note the log-scale of the y-axis.
}
\label{fig:Dip0}
\end{figure*}
Hence,
excitation with thermal light is completely different than that of pulses due to its broad spectrum,
which simultaneously excites  the four chromophores in accord with
the magnitude of the  transition dipole moment (see Table \ref{tab:paramPC645}).
For example, excitation of the MBV molecules in Fig.~\ref{fig:sitebasisnatural}
is due to both direct excitation by the light as well as energy transfer.
Figure~\ref{fig:Dip0} quantifies each contribution by showing
the dynamics of the populations when the coupling of the
MBV molecules  to the light is turned off for $\Lambda=0$ (left panel) and $\Lambda=13$~cm$^{-1}$
(right panel).
For $\Lambda=0$, at 2~ps,  the population of the MBV molecules is ca. three orders of 
magnitude smaller than the population 
obtained when the direct interaction with light is 
included (see Fig.~\ref{fig:sitebasisnatural}).
For $\Lambda=13$~cm$^{-1}$, a significant increment in the population of MBV molecules 
is achieved, ca. two orders of magnitude.
This points out the important role of vibrations in exciton energy transfer even under incoherent
excitation. This is made even all the more evident when examining the case
of $\Lambda=130~\mathrm{cm}^{-1}$, a value suggested experimentally for the PC645
system and shown in Fig. ~\ref{fig:Lambda130}. Here (by comparison with 
extrapolated results from Fig. \ref{fig:sitebasisnatural}) excited 
MBV population arising from direct MBV excitation and from DBV to MBV transfer
are seen to be comparable in size.

\begin{figure*}[h]
\includegraphics[width=0.49\columnwidth]{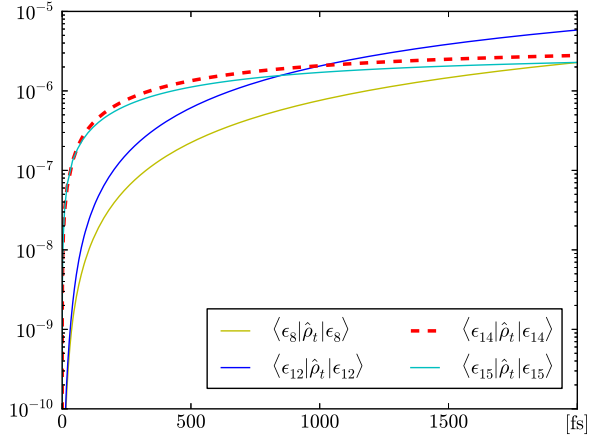}
\vspace{-0.5cm}
\caption{Time evolution the states  associated with the single exciton manifold in the site basis
with no coupling of the MBV molecules to the thermal light  and  with 
$\Lambda=130$~cm$^{-1}$.
Note the log-scale of the y-axis.
}
\label{fig:Lambda130}
\end{figure*}

%

\subsection{The Requirement for Proper Preparation}
Several studies have been published that ignore the preparation step, where the system is 
excited by light,
and consider that the system starts in a given excited state, after which the dynamics follows.
For example, in the PC645 case such a study   \cite{FOS12} assumed that the sunlight had prepared the 
system in eigenstates where the excitations of DBVc and DBVd are mainly resident.
The subsequent effect of sunlight is included via a master equation of the
Lindblad form with a decay rate $\Gamma_{\mathrm{Lindblad}}^{-1} \sim 250$~ps.
Our approach allows us to assess the relevance of this
preparation step to the natural light-induced process.

\begin{figure}[h]
 \includegraphics[width=0.49\columnwidth]{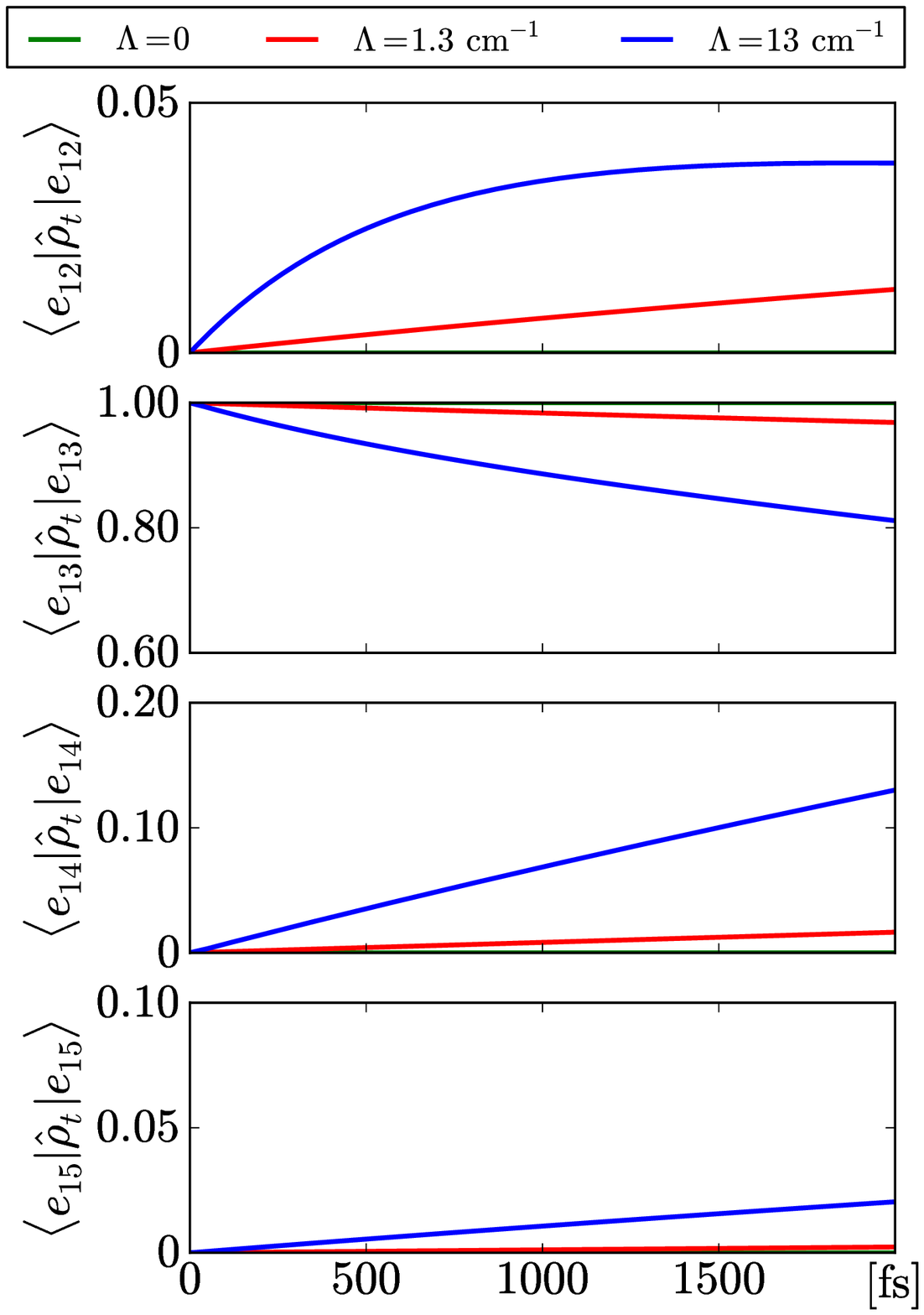}
\includegraphics[width=0.49\columnwidth]{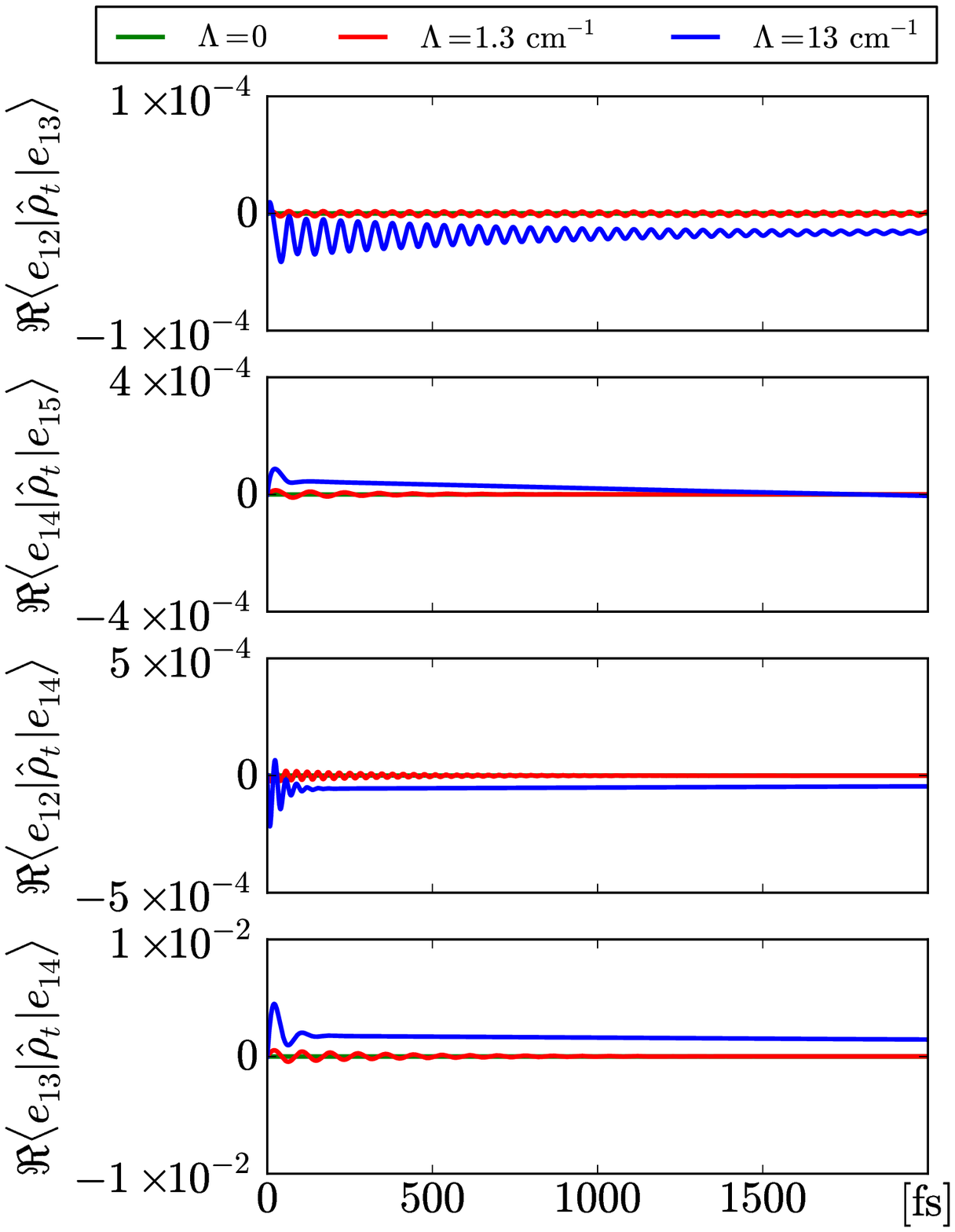}
\vspace{-0.5cm}
\caption{
Left panel: Time evolution at room temperature of the states associated with the single exciton manifold for different $\Lambda$ values.
Right panel: Time evolution at room temperature  of several of the superpositions between
single-exciton-manifold states induced by sunlight.
Parameters are $\lambda = 100$~cm$^{-1}$, $T_{\mathrm{env}} = 300$~K, $T_{\mathrm{bb}} = 5600$~K
and $\hat{\rho}(t=0)=\proj{e_{13}}{e_{13}}$.
}
\label{fig:excbasisartificial}
\end{figure}

\begin{figure}[h]
 \includegraphics[width=0.49\columnwidth]{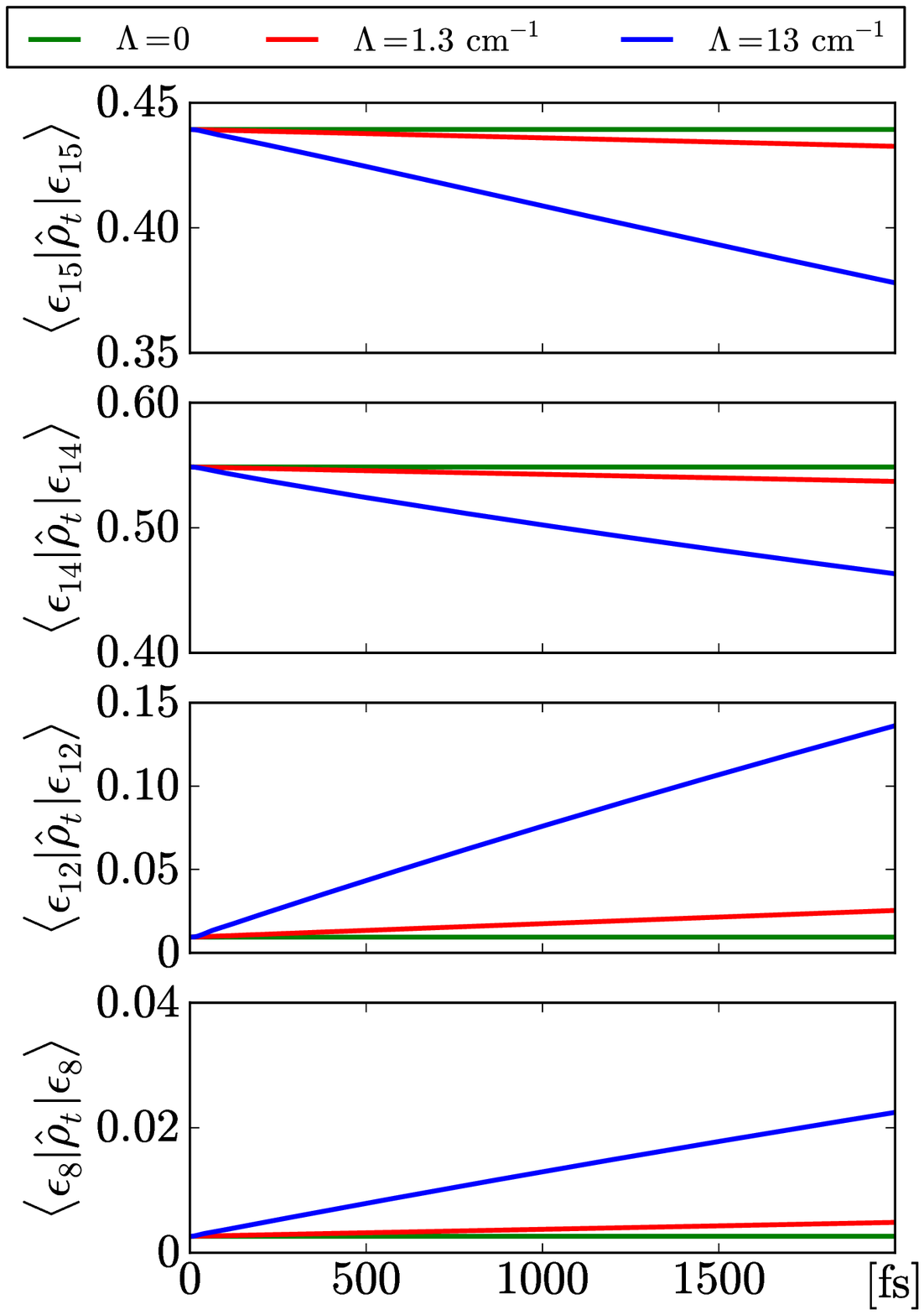}
\includegraphics[width=0.49\columnwidth]{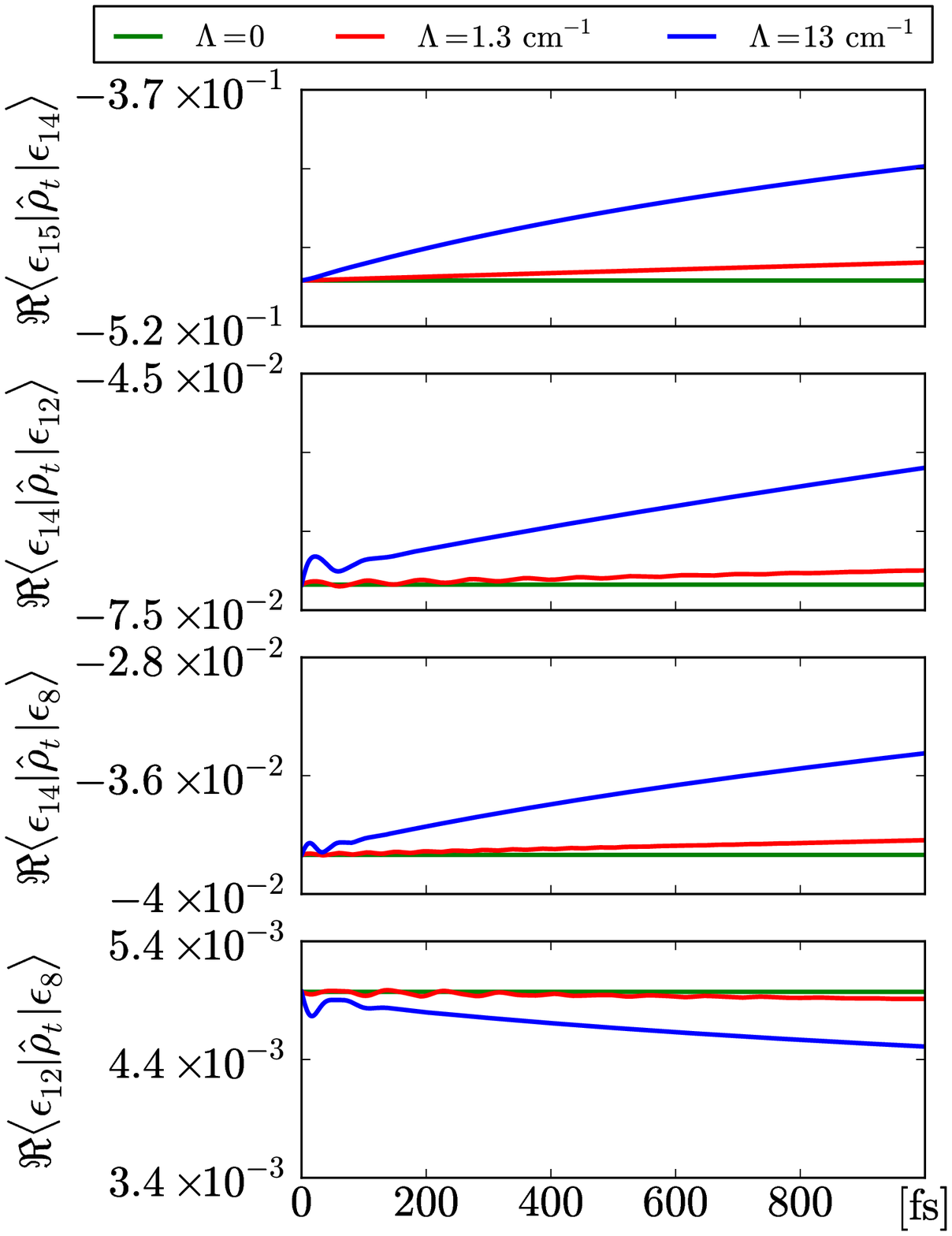}
\vspace{-0.5cm}
\caption{Description and parameters as in Fig.~\ref{fig:excbasisartificial} but with matrix
elements are in the site basis. }
\label{fig:sitebasisartificial}
\end{figure}

 Consider then the case introduced in \cite{FOS12} where the antenna is
artificially prepared in the excited eigenstate $\proj{e_{13}}{e_{13}}$ and the interaction with sunlight
and the local environment is then switched on, defining $t=0$.
Figures \ref{fig:excbasisartificial} and \ref{fig:sitebasisartificial} show the resultant
dynamics in the eigenstate and the site basis, respectively.
Note first that the amplitude of the population and the coherences are approximately five orders of magnitude
larger here than in the case of natural excitation.
This implies that the artificial initial condition
$\hat{\rho}(t=0)=\proj{e_{13}}{e_{13}}$ strongly activates the subsequent interaction with  the environment.
That is, under these  conditions the dynamics is dominated by the environment, with marginal influence of the radiation.
To quantify this statement, we calculate the trace distance
$\mathcal{D}(\hat{\rho},\hat{\rho}_{\hat{\boldsymbol{\mu}}=0}) = \mathrm{tr}|\hat{\rho}-\hat{\rho}_{\hat{\boldsymbol{\mu}}=0}|$
between the density operator driven by the (environment + sunlight) $\hat{\rho}$, and the density
operator driven only by the environment, $\hat{\rho}_{\boldsymbol{\mu}=0}$.
For orthogonal states, i.e.,  completely distinguishable states, the trace distance would be one.
For the cases depicted in Figs.~\ref{fig:excbasisartificial} and \ref{fig:sitebasisartificial}, the trace
distance is of the order of $10^{-15}$.
This clearly demonstrates  that the effect of the radiation is negligible in this case.
That is, the coherences observed in
Fig.~\ref{fig:excbasisartificial} have nothing to do with sunlight but are generated
by the non-equilibrium initial condition \cite{PB13}, contrary to an earlier incorrect
assertion \cite{FOS12}.
It also emphasizes the need to properly include the radiative excitation step to study
dynamics induced by incoherent light,  as done in this paper.

\section{Summary}
An open systems quantum perspective is adopted to examine the excitation, via incoherent light,
of systems imbedded in a bath. 
Specifically, the nature and role of the quantum dynamical maps are analyzed in detail for model 
components of light harvesting systems.
Contrasts between the effect of the radiative bath and the phonon bath are analyzed.
In particular,
phonon-bath dephasing times are found to be far longer in the radiative case than in the phonon case
due to well described differences in system-bath coupling and bath spectral densities. Examination
of increasingly complicated models of components of the PC645 light harvesting complex shows
the appearance of oscillatory and stationary coherences, the significance of multichromophoric
excitation that occurs naturally as distinct from localized chromophore excitation that is
carried out in the laboratory, and the
importance of properly treating the light-excitation step to model the true dynamics. The approach
provides a direct means of analyzing a wide variety of initial conditions within one formalism.

%
%
%
%
%
%
%
%
%

\textit{Acknowledgements}\textemdash
This work was supported by the US Air Force Office of Scientific Research under contract
number FA9550-13-1-0005, by \textit{Comit\'e para el Desarrollo de la Investigaci\'on}
(CODI) of Universidad de Antioquia, Colombia under the \textit{Estrategia de Sostenibilidad}
and the \textit{Purdue-UdeA Seed Grant Program} and by the \textit{Departamento Administrativo
de Ciencia, Tecnolog\'ia e Innovaci\'on} (COLCIENCIAS) of Colombia under the contract
number 111556934912.

\bibliography{drsprlv3}

\newpage
\appendix
\section{Slow Turn-On of the Light}
To simulate the slow turn-on of the light, it is convenient to introduce a time dependence  of the
transition dipole moments, namely, $\boldsymbol{\mu}_{ab} = \boldsymbol{\mu}_{ab} \mathrm{Erf}_{ab}(t) $,
with
\begin{equation}
\mathrm{Erf}_{12,13}(t) = \frac{2}{\sqrt{\pi}} \int_0^{ |\omega_{12,13}| t/\alpha} \mathrm{d} z \exp(-z^2).
\end{equation}
For large $\alpha\gg1$ this can be approximated by $\mathrm{Erf}_{12,13}(t) \sim \frac{2|\omega_{12,13}|}{\pi \alpha} t$.
Thus, $2|\omega_{12,13}|/\pi \alpha$ can be interpreted as the rate of light turning-on. The method
introduces the turn-on rate into the dipole coupling and is
different from that advanced in Ref. \cite{DTB16b}, where the field populations are altered,
but should give  similar results.

\begin{figure*}[h]
\includegraphics[width=0.49\columnwidth]{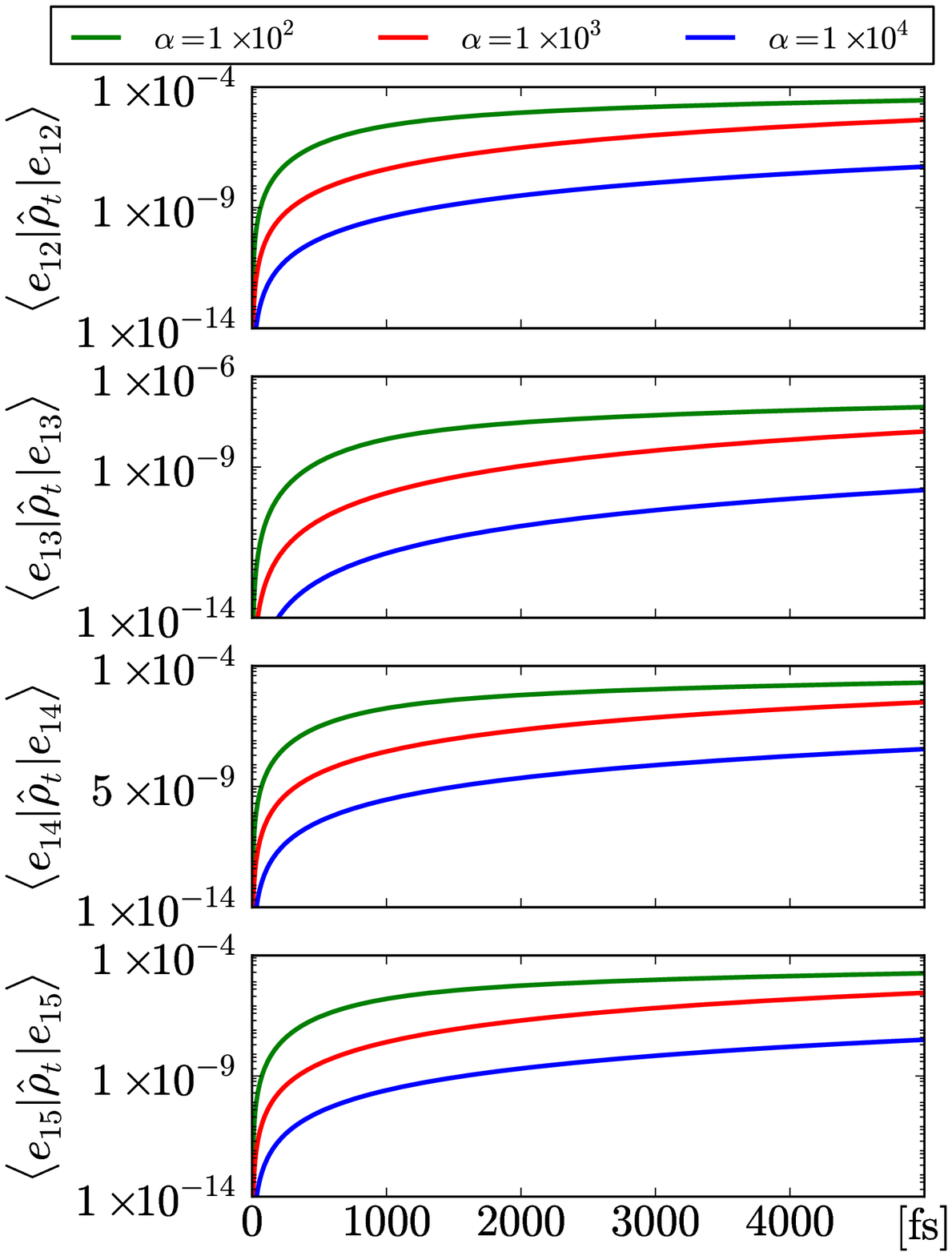}
\includegraphics[width=0.49\columnwidth]{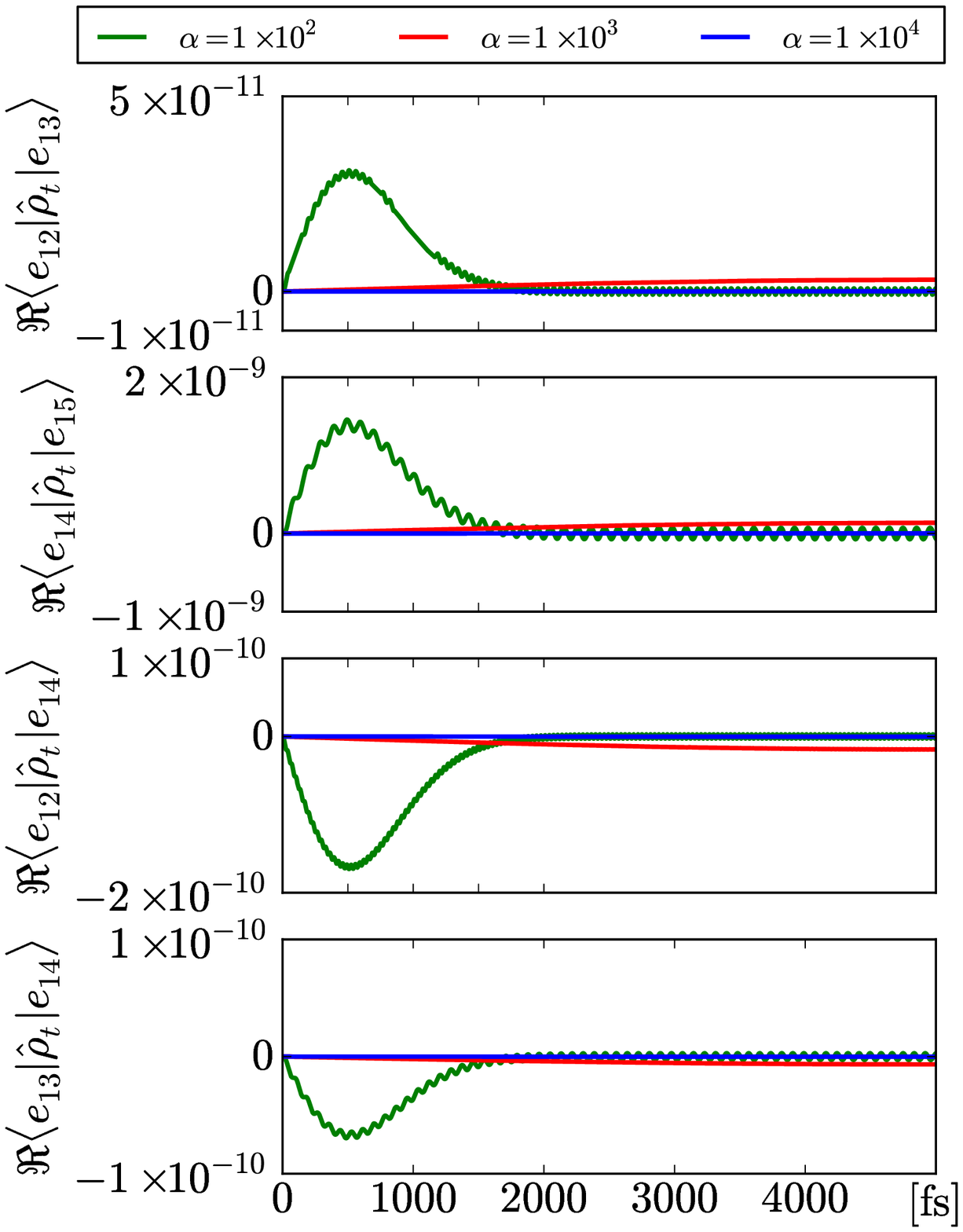}
\vspace{-0.5cm}
\caption{Time evolution at room temperature of the populations (left panels) and coherences (right panels)
associated with the single exciton manifold for different values of $\alpha$ with $\Lambda =0$.
Parameters as in Fig.~\ref{fig:excbasisnatural}.}
\label{fig:SlowTurnOn}
\end{figure*}
Figure~\ref{fig:SlowTurnOn} depicts the time evolution of the single exciton manifold of the PC645 sub-unit 
discussed above.
Clearly, the slower the turn-on the smaller the coherences.

\end{document}